\documentclass[reprint,superscriptaddress]{revtex4-1}
\usepackage{url}
\usepackage{amsmath}
\usepackage{mathtools}
\usepackage{bbold}
\usepackage{mathrsfs}

\usepackage{graphicx}
\usepackage[caption=false]{subfig}
\usepackage{epstopdf} 
\usepackage{wrapfig}
\usepackage{braket}
\usepackage{xcolor}
\usepackage{comment}
\usepackage{lipsum}
\usepackage{cancel}
\usepackage{hyperref}

\newcommand{\RNum}[1]{\uppercase\expandafter{\romannumeral #1\relax}}

\newcommand{\btheta}{\boldsymbol{\theta}}

\def\ket#1{| #1\rangle}

\def\ud{\mathrm{d}}

\newcommand{\nep}{\mathrm{e}}

\newcommand{\defuguale}{\stackrel{\mathrm{def}}{=}}

\newcommand{\LO}{\mathrm{\scriptscriptstyle LO}}
\newcommand{\dQA}{\mathrm{\scriptscriptstyle dQA}}
\newcommand{\QAOA}{\mathrm{\scriptscriptstyle QAOA}}

\newcommand{\Real}{\mathrm{Re}}

\newcommand{\C}{{\mathbf C}}
\newcommand{\rmC}{{\mathrm C}}

\newcommand{\target}{\mathrm{\scriptscriptstyle targ}}
\newcommand{\drive}{\mathrm{\scriptscriptstyle drive}}

\newcommand{\best}{\mathrm{\scriptstyle B}}
\newcommand{\Nbasis}{\mathrm{N_c}}

\newcommand{\PauliSigma}{\hat{\sigma}}

\newcommand{\Ho}{\hat{H}}
\newcommand{\Uo}{\hat{U}}

\newcommand{\res}{\mathrm{res}}
\newcommand{\fin}{\mathrm{fin}}
\newcommand{\eff}{\mathrm{eff}}

\newcommand{\Ptrot}{\mathrm{P}}

\graphicspath{{./}{./Fig/}}

\begin{document}

\author{Giovanni Pecci}
\affiliation{CNR-IOM - Istituto Officina dei Materiali, Consiglio Nazionale delle Ricerche, c/o SISSA Via Bonomea 265, 34136 Trieste, Italy}

\author{Ruiyi Wang}
\affiliation{SISSA, Via Bonomea 265, I-34136 Trieste, Italy}

\author{Pietro Torta}
\affiliation{SISSA, Via Bonomea 265, I-34136 Trieste, Italy}
\affiliation{Dipartimento di Fisica, Università degli Studi di Milano, Via Celoria 16, 20133 Milano, Italy}

\author{Glen Bigan Mbeng}
\affiliation{Universit\"at Innsbruck, Technikerstraße 21 a, A-6020 Innsbruck, Austria}
\affiliation{Parity Quantum Computing GmbH, A-6020 Innsbruck, Austria}

\author{Giuseppe Santoro}
\affiliation{CNR-IOM - Istituto Officina dei Materiali, Consiglio Nazionale delle Ricerche, c/o SISSA Via Bonomea 265, 34136 Trieste, Italy}
\affiliation{SISSA, Via Bonomea 265, I-34136 Trieste, Italy}
\affiliation{International Centre for Theoretical Physics (ICTP), P.O.Box 586, I-34014 Trieste, Italy}

\title{Beyond Quantum Annealing: Optimal control solutions to MaxCut problems}

\begin{abstract} 
Quantum Annealing (QA) relies on mixing two Hamiltonian terms, a simple driver and a complex problem Hamiltonian, in a linear combination. 
The time-dependent schedule for this mixing is often taken to be linear in time: improving on this linear choice is known to be essential and has proven to be difficult.
Here, we present different techniques for improving on the linear-schedule QA along two directions, conceptually distinct but leading to similar outcomes: 1) the first approach consists of constructing a Trotter-digitized QA (dQA) 
with schedules parameterized in terms of Fourier modes or Chebyshev polynomials, inspired by the Chopped Random Basis algorithm (CRAB) for optimal control in continuous time; 
2) the second approach is technically a Quantum Approximate Optimization Algorithm (QAOA), whose solutions are found iteratively using linear interpolation or expansion in Fourier modes. 
Both approaches emphasize finding smooth optimal schedule parameters, ultimately leading to hybrid quantum-classical variational algorithms of the alternating Hamiltonian Ansatz type. 
We apply these techniques to MaxCut problems on weighted 3-regular graphs with $N=14$ sites, focusing on hard instances that exhibit a small spectral gap, for which a standard linear-schedule QA performs poorly. 
We characterize the physics behind the optimal protocols for both the dQA and QAOA approaches, discovering \textit{shortcuts to adiabaticity}-like dynamics.  
Furthermore, we study the transferability of such smooth solutions among hard instances of MaxCut at different circuit depths. 
Finally, we show that the smoothness pattern of these protocols obtained in a digital setting enables us to adapt them to continuous-time evolution, contrarily to generic non-smooth solutions. This procedure results in an optimized quantum annealing schedule that is implementable on analog devices.
\end{abstract}

\maketitle

\section{Introduction}
The steady advances of quantum computing and quantum technologies sparked great interest in exploring applications to problems that might be out-of-reach for classical devices~\cite{preskill2023quantum}. 
This interest extends across various fields, including the simulation of quantum systems, quantum chemistry, and classical optimization problems~\cite{abbas2023quantum, 10.3389/fphy.2014.00005}. 
While state-of-the-art quantum devices may still lack the required computational power to surpass their classical counterparts and provide ground-breaking contributions to these interdisciplinary problems
\cite{guerreschi_qaoa_2019, Preskill2018quantumcomputingin},
it is crucial to carefully design quantum algorithms to optimally exploit the available computational resources. 

Quantum Annealing (QA)~\cite{finnila_quantum_1994,Santoro_SCI02,Santoro_2006,Albash_RMP18} offers an approach for addressing these challenges by exploiting adiabatic dynamics. 
In the standard framework, an interpolating Hamiltonian $\Ho(t) = (t/\tau) \Ho_{\target} + (1-t/\tau) \Ho_{\drive}$ is constructed, and the (allegedly) unitary Schr\"odinger evolution of the state $|\psi(t)\rangle$ is followed from time $t=0$ --- where the ground state $|\psi_0\rangle$ of $\Ho_{\drive}$ is easy to prepare --- to some suitably large annealing time $\tau$, with the goal of reaching the
target ground state $|\psi_{\target}\rangle$ of $\Ho_{\target}$.

In this flavor, which we shall dub {\em linear-schedule} QA, it remains unclear whether actual quantum speed-up~\cite{q_speedup} should be expected. 
A notable example is provided by the provable quadratic speedup for the Grover search problem~\cite{Grover_PRL1997,Zalka_PRA1999}.
As pointed out long ago by Roland \& Cerf \cite{Roland_PRA2002}, a linear-schedule QA of the Grover problem would suffer from the instantaneous gap $\Delta$ closing as $\Delta\sim 1/\sqrt{N}$ --- where $N=2^n$ is the Hilbert space dimension for $n$ qubits; this fact, together with the standard Landau-Zener \cite{Zener_PRS32} mechanism predicting that $\tau\propto 1/\Delta^2$, results in a total annealing time scaling as $\tau\propto N$.
On the contrary, an appropriate tailoring of the schedule $s(t)$ in the interpolating Hamiltonian 
$\Ho(s(t)) = s(t) \Ho_{\target} + (1-s(t)) \Ho_{\drive}$ results in the quadratic Grover speed-up, with $\tau\propto 1/\Delta = \sqrt{N}$.
This example clearly shows that an improvement over a linear schedule $s(t)=t/\tau$ can be essential to achieve quantum speed-up, whenever possible. 
How to do that is a complicated issue, which we will partly address in the present paper. 

On a broader perspective, there is a large body of literature showing that QA faces strong limitations in the presence of exponentially small energy gaps, which can result in exponentially increasing annealing times with system size~\cite{Caneva_PRB2007,Knysh_NatComm16,Knysh_PRA2020,Zamponi_QA:review}.
Various techniques have been proposed to overcome this limitation within the paradigm of continuous-time Schr\"odinger evolution, for instance employing the toolbox of Quantum Control (QC)~\cite{Dalessandro2007}.

A general strategy would be to perform a state-dependent Shortcut to Adiabaticity (STA)~\cite{STA_Review_RMP2019}, literally a ``fast route'' to the ground state of the final target Hamiltonian in a shorter time $\tau$, as compared to a slow QA (adiabatic) evolution. 
A particular STA technique, known as transitionless or counter-diabatic driving~\cite{Rice_JPCA2003,Berry_2009}, involves adding particular terms to the time-evolving Hamiltonian, which are designed to suppress transitions out of the instantaneous ground state. However, exact implementations of this scheme are difficult in many-body systems, as they require hard-to-devise non-local terms. Variational approaches have been introduced~\cite{KOLODRUBETZ20171, Matsuura_PRA2021, Susa_PRA2021, barraza_2023variational} to overcome these difficulties. 

Other techniques have been specifically introduced within the field of 
Adiabatic Quantum Computation~\cite{Albash_RMP18}, 
including the addition of waiting times (pauses) in the schedule~\cite{Passarelli_PRB2019}, 
reverse quantum annealing~\cite{Passarelli_PRA2020}, or 
the introduction of Hamiltonian terms that transform the nature of the critical points encountered during the evolution, from 1$^{\textrm{st}}$ to 2$^{\textrm{nd}}$ order~\cite{Seoane_JPA2012}.

Various methods of Quantum Control appear particularly well-suited for achieving a ground-state STA, notably the Chopped RAndom Basis (CRAB) algorithm~\cite{Caneva_PRA2011, Doria_PRL2011} and follow-up developments~\cite{Montangero_dCRAB_PRA2015,koch_quantum_2022}.
In CRAB, the various control functions are expanded in terms of a finite basis set of functions, usually a Fourier basis --- but polynomials have also been used~\cite{Quiroz_PRA2019} ---,
hence transforming the functional minimization problem into a finite-dimensional optimization. 
Alternatively, a linear piece-wise decomposition of the schedule between a series of randomly chosen points has been explored~\cite{Côté_2023} to optimize a frustrated Ising ring model~\cite{Knysh_PRA2020}, which shows an exponentially small gap in the excitation spectrum, thereby posing severe difficulties for QA. 

Parallel developments in the field of Variational Quantum Algorithms (VQA) have approached the problem from the alternative perspective of a digitized quantum evolution~\cite{Nature_dQA}, whereby a series of different unitary operators are applied to the initial simple-to-construct state $|\psi_0\rangle$. 
These approaches include the Quantum Approximate Optimization Algorithm (QAOA)~\cite{Farhi_arXiv2014}, 
originally introduced as an alternative to QA for classical optimization problems, 
and the Variational Quantum Eigensolver (VQE)~\cite{Tilly_VQE_PhysRep2022}, 
specifically aiming at solving quantum chemistry problems and, more generally, quantum ground state preparation.
These digitized techniques offer the practical advantage of a native digital setting that is well adapted to universal 
gate-based~\cite{Nielsen_Chuang:book} quantum computing hardware, and encompass a large class of hybrid quantum-classical algorithms 
~\cite{cerezo_variational_2021, NISQ_Algorithms_RMP2022, lloyd2018quantum}.  

Yet, the trade-off between the expressivity and the trainability of parameterized quantum circuits remains a substantial challenge for the practical implementation of this framework on real quantum 
devices~\cite{cerezo2023does, Bittel_PRL2021}. 
On top of the presence of statistical and physical noise, the optimization landscape of non trivial (or classically-simulable) parameterized quantum circuits is affected by the presence of low-quality local minima~\cite{Bittel_PRL2021} and vanishing gradients (barren plateaus)~\cite{McClean_2018, ragone2023unified, fontana2023adjoint}.
Several approaches to limit or avoid barren plateaus have been proposed 
and applied to specific tasks~\cite{MPS_QAOA_BP, Skolik_2021, INITbarrenGrant_2019, Kliesch_2021, BPshadows, Mele_PRA2022, Mathey_PRR2024}.
At the same time, developing smart initializations~\cite{MPS_pre_training} or iterative optimization schemes~\cite{zhou_quantum_2020} may prove essential to successfully train a VQA.

Quantum optimal control techniques may help, in this digital framework, to single out optimal or nearly-optimal variational parameters that are smooth functions of the layer index, which have been observed in several works~\cite{mbeng_quantum_2019,zhou_quantum_2020,Sack2021quantumannealing, Wauters_PRR2020, Torta_PRB2023, Wurtz2022counterdiabaticity,Pagano_PNAS2020, LGT_QAOA_2021}. 
Such smooth optimal solutions exhibit notable properties and have proved effective in avoiding local minima~\cite{zhou_quantum_2020} and circumventing barren plateaus~\cite{Mele_PRA2022}. %
Remarkably, it has been empirically demonstrated~\cite{Pagano_PNAS2020, Mele_PRA2022} that this class of solutions is transferable across different problem sizes, from small to large, for the quantum many-body ground state preparation at any point of the phase diagram, highlighting their versatility and practical applicability in hybrid quantum-classical computing.

In this paper, we utilize established quantum optimal control methods to systematically generate smooth optimal solutions for combinatorial optimization problems. 
For definiteness, our focus is on MaxCut instances of weighted 3-regular graphs, but our techniques are straightforwardly generalized to different problems.
In particular, we select hard MaxCut instances characterized by a small energy gap, a regime in which standard {\em linear-schedule} QA is expected to yield poor performance.
The rationale behind this choice is that a closing gap is expected to occur in most real-world use cases of NP-hard classical optimization problems tackled with QA, especially for growing system size~\cite{Knysh_NatComm16}.

We characterize the dynamics implemented by such smooth and optimal parameters, unveiling the non-adiabatic mechanism that enables these techniques to outperform adiabatic protocols. 
Additionally, we observe promising transferability results across different \emph{hard} instances of the problem. 
Although these techniques are applied in a discrete-time setting, making them suitable for digital devices, we demonstrate that smooth solutions can be effectively transferred to analog devices with comparable performance. 
This highlights the adaptability of smooth solutions for continuous-time quantum annealing, yielding an optimized problem-dependent schedule $s(t)$, thereby broadening their range of applicability across different quantum computing platforms.

The paper is structured as follows. 
In Section \ref{sec:methods} we present the different methods we implemented to generate smooth solutions. In Section \ref{sec:maxcut} we describe the MaxCut problem and the procedure utilized to single out hard instances. In Section \ref{sec:results} we present our results, and we draw a comparison among the different techniques and their performance. 
We characterize the non-adiabatic mechanism associated with smooth solutions and prove transferability among different hard instances. 
Finally, we implement these smooth solutions in a continuous-time evolution framework, showing their suitability for analog quantum devices. 
In Sec. \ref{sec:conclusions} we summarize the significance of our results and discuss possible future work.

\section{From Quantum Annealing to Variational Quantum Algorithms} \label{sec:methods}

\subsection{Quantum Annealing and optimal control} \label{sec:QA}
This section provides a concise summary of Quantum Annealing (QA) and quantum optimal control, focusing on a digitized version of the quantum dynamics.

In the standard formulation of QA, also implemented on analog quantum devices~\cite{Johnson_Nat11}, the interpolating Hamiltonian is $\Ho(s(t)) = s(t) \Ho_{\target} + (1-s(t)) \Ho_x$.
Here, $\Ho_{\target}$ represents the target (or problem) Hamiltonian, whose ground state is the desired outcome — typically formulated using spin-$1/2$ Pauli operators. 
Conversely, the driver $\Ho_x$ acts as a non-commuting operator that introduces quantum fluctuations, often exemplified by a transverse field term of the form $\Ho_x=- \Gamma_0 \sum_j \PauliSigma^x_j$.
Ideally, QA relies on an adiabatic Schr\"odinger dynamics~\cite{Albash_RMP18} by slowly increasing $s(t)$ from $s(0)=0$ to $s(\tau)=1$ in a large total annealing time $\tau$. This dynamics starts from the trivial ground state of $\Ho_x$, which we dub $ \ket{\Psi_0}$, and following the instantaneous ground state of $\Ho(s)$ would eventually lead to the target ground state. However, a critical challenge arises from the necessity of exceedingly large annealing times $\tau$ that are required to adiabatically follow the instantaneous ground state of $\Ho(s)$, especially when the system crosses a critical point or a first-order phase transition~\cite{Zamponi_QA:review} at some intermediate value $s_c$, which is generally unknown.
Adapting the schedule $s(t)$ so as to {\em slow down} in the proximity of $s_c$ is also generally unfeasible~\cite{Cubitt_Nat2015}, 
so that one usually adopts a {\em linear} schedule $s(t)=t/\tau$, pushing $\tau$ to large values. 

An {\em optimal control problem} can be similarly formulated. 
Let us assume, for definiteness, that we are dealing with a classical optimization problem, mapped onto a diagonal $\Ho_{\target}=\Ho_z$, which is simply built from $\PauliSigma^z$ operators.
Later on, we shall restrict our numerics to MaxCut.  
The system follows a unitary Schr\"odinger dynamics 
\begin{equation}
i\hbar \frac{\ud}{\ud t} |\Psi(t)\rangle = \Ho(t) |\Psi(t)\rangle \;,
\label{eqn:Schrodinger_eq}
\end{equation}
with a time-dependent Hamiltonian of the interpolating form
\begin{equation}
     \Ho(t) = A^z(t) \, \Ho_z + A^x(t) \, \Ho_x \;,
\label{eqn:ann_hamiltonian}
\end{equation}
where we impose the boundary conditions $A^z(0)=0$, $A^x(\tau)=0$, and $A^z(\tau)>0$.
The goal is to determine the optimal form of the two driving fields $A^z(t)$ and  $A^x(t)$ such that the variational energy
\begin{equation}
E^{\fin}(\tau) = \langle \Psi(\tau) | \Ho_z | \Psi(\tau)\rangle 
\label{eqn:final_avg_energy}
\end{equation}
is {\em minimal}.
This optimal control problem can be formulated either for a predetermined total time $\tau$, or by including the optimization of $\tau$ itself as part of the task.

Such a functional minimization is generally intractable.
To simplify it, we transform it into a {\em finite-dimensional} minimization by using the so-called Chopped RAndom Basis (CRAB)  algorithm~\cite{Muller_2022, Doria_PRL2011, Montangero_PRA2011, Montangero_PRA2015}, 
i.e., we expand the fields in terms of a selected discrete set of $\Nbasis$ basis functions $\{f_n(t)\}$:
\begin{equation}
\left\{
\begin{array}{l} 
A^x(t) = \displaystyle \Big(1-\frac{t}{\tau}\Big) \Big( 1 + \sum_{n=1}^{\Nbasis} \rmC^x_n\, f_n(t) \Big) \vspace{3mm} \\
A^z(t) = \displaystyle \frac{t}{\tau} \Big( 1 + \sum_{n=1}^{\Nbasis} \rmC^z_n\, f_n(t) \Big) 
\end{array}
\right. \;.
\label{eqn:CRAB_decomposition}
\end{equation}
Notice that the fields automatically satisfy the two constraints $A^z(0)=0$ and $A^x(\tau)=0$, and reduce to the linear schedule QA form, i.e.\
$A^z(t)=s(t) = t/\tau$ and $A^x(t)=1-s(t)$,  
if all the expansion coefficients $C^{x/z}_n$ are set to zero. 
Concerning the basis functions, a common choice is given by Fourier modes. Here we take:
\begin{equation}
f_n(t) = \sin(\omega_n t) \hspace{10mm} \omega_n = \frac{\pi n}{\tau} (1+r_n) \;,
\end{equation}
where $r_n\in [-\frac{1}{2},\frac{1}{2}]$ is a random number, introducing a multiplicative noise in the frequencies, which is useful to improve the expressive power of this finite-dimensional {\em Ansatz}~\cite{Montangero_PRA2011}. 
We dub this choice F-CRAB, where F stands for Fourier.
We have verified empirically that this works better than an additive noise $\omega_n = (\pi/\tau) (n+r_n)$.

An alternative option consists of using 
\begin{equation}
f_n(t) = \cos\Big( n (1+r_n) \arccos(t/\tau)\Big) \;,
\end{equation}
where $r_n\in [-\frac{1}{2},\frac{1}{2}]$  is again a random number.
For $r_n=0$, these functions coincide with the Chebyshev polynomials of the first kind $T_n(t/\tau)$, while for $r_n\neq 0$ they are natural extensions of polynomials 
with non-integer powers~\cite{boyd2004convex,duffin_geometric_1973,MARANAS1997351}, often known as signomials.
We refer to this choice as C-CRAB.

In both variants of the CRAB algorithm, the variational energy in Eq.~\eqref{eqn:final_avg_energy} is optimized with respect to the set of $2 \Nbasis$ expansion coefficients $C^{x/z}_n$. Instead of repeatedly performing multiple local optimizations on the same energy landscape from various starting points, the strategy involves generating a new landscape each time. This is achieved by selecting a different set of random frequencies~\cite{Montangero_PRA2011}.

To classically simulate the quantum dynamics in Eq.~\ref{eqn:Schrodinger_eq}, a time-discretization is necessary: the continuous-time dynamics is discretized by introducing a finite --- in principle, sufficiently small --- time-step, denoted as $\Delta_t=\tau/\Ptrot$. 
We define an associated time grid $t_m=(m-\frac{1}{2}) \, \Delta_t$, with $m=1\cdots\Ptrot$, at which we evaluate the fields, setting $A^{x/z}_m=A^{x/z}(t_m)$. 
We then approximate the fields as constant in each interval 
$[t_m-\Delta_t/2,t_m+\Delta_t/2]$.
The ensuing evolution operator over a time-step $\Delta_t$ is further approximated with a first-order Trotter splitting:
\begin{equation} \label{eqn:1st_Trotter}
\nep^{-i (\theta^x_m \Ho_x + \theta^z_m \Ho_z)} \approx 
\nep^{-i\theta^x_m \Ho_x} 
\nep^{-i\theta^z_m \Ho_z} \defuguale \Uo(\theta^x_m,\theta^z_m) \;,
\end{equation}
where we defined (by setting $\hbar=1$)
\begin{equation} \label{eq:theta_x_z}
\left\{
\begin{array}{l}
\theta^x_m = \Delta_t A^x_m  
= {\textstyle\frac{\Ptrot-(m-\frac{1}{2})}{\Ptrot}} \Delta_t
\Big(1+{\displaystyle\sum_{n=1}^{\Nbasis}} \rmC^x_n f_n(t_m)\Big)  \vspace{4mm} \\
\theta^z_m = \Delta_t A^z_m = 
{\textstyle\frac{(m-\frac{1}{2})}{\Ptrot}} \Delta_t
\Big(1+{\displaystyle\sum_{n=1}^{\Nbasis}} \rmC^z_n f_n(t_m)\Big)
\end{array}
\right. \;.
\end{equation}

The time-discretized and Trotter-split Schr\"odinger evolution results in a \emph{digitized} dynamics leading to the final state
\begin{equation} \label{eqn:dQA_state}
|\Psi_{\Ptrot}^{\dQA}(\btheta)\rangle = \Uo(\theta^x_\Ptrot,\theta^z_\Ptrot) \, 
\cdots 
\Uo(\theta^x_1,\theta^z_1) \, |\Psi_0\rangle \;,
\end{equation}
where $\btheta=(\btheta^x,\btheta^z)$. 
Note that $\btheta^x=(\theta^x_1\cdots\theta^x_{\Ptrot})$ and 
$\btheta^z=(\theta^z_1\cdots\theta^z_{\Ptrot})$ are both 
$\Ptrot$-dimensional vectors, each depending on the $\Nbasis$ expansion coefficients $\C^x=(\rmC^x_1\cdots\rmC^x_\Nbasis)$ and $\C^z=(\rmC^z_1\cdots\rmC^z_\Nbasis)$, respectively.
Albeit we denote this state with the apex dQA, an acronym for digitized Quantum Annealing~\cite{Nature_dQA,Mbeng_dQA_PRB2019}, this is actually a digitized version of a generic optimal control dynamics, reducing to the linear-schedule digitized-QA of Ref.~\cite{Mbeng_dQA_PRB2019} only when all the expansion coefficients $C^{x/z}_n$ vanish.

Previous evidence on linear-schedule dQA~\cite{Mbeng_dQA_PRB2019} showed that even a first-order Trotter splitting is adequate up to values of $\Delta_t$ of order $1$.
In principle, the order of the Trotter approximation could be increased by utilizing, e.g., a second or fourth-order expansion. 
However, the focus here relies on the optimization of the variational parameters in the digitized state rather than on approximating a prescribed continuous-time evolution.
Therefore, we keep the form of the state in 
Eqs.~\eqref{eqn:1st_Trotter}-\eqref{eqn:dQA_state}, 
since improving the Trotter order would lead to the same alternating {\em Ansatz} structure~\cite{QAlternatingOA}, only at a higher depth.

Finally, to enlarge the variational manifold, it would be beneficial to include the time-step $\Delta_t$ as an additional variational parameter.
To keep a \emph{linear} dependence of the angles in Eq.~\ref{eq:theta_x_z} on the set of variational parameters, we rather fix $\Delta_t=1$, and introduce a new free parameter $\rmC_0$ as follows:
\begin{equation} \label{eqn:theta_vs_C}
\left\{
\begin{array}{l}
\theta^x_m  = 
\displaystyle \frac{\Ptrot-(m-\frac{1}{2})}{\Ptrot} \Big(\rmC_0 + \sum_{n=1}^{\Nbasis} \rmC^x_n f_n(t_m)\Big) \vspace{5mm} \\
\theta^z_m = 
\displaystyle \frac{(m-\frac{1}{2})}{\Ptrot} \Big(\rmC_0 + \sum_{n=1}^{\Nbasis} \rmC^z_n f_n(t_m)\Big) 
\end{array}
\right. \;.
\end{equation}
This approach leads to a digitized version of quantum optimal control without a constraint on the time-step $\Delta_t$, while preserving the linear dependence of the angles on the variational parameters.
Indeed, an optimal protocol characterized by the ${2\Nbasis+1}$ variational parameters $\C=(\rmC_0,\C^x,\C^z)$ is equivalent to the digitized state with angles in Eq.~\ref{eq:theta_x_z}, by identifying $\rmC_0$ with $\Delta t$ and rescaling the other ${2\Nbasis}$ parameters by its value.

\begin{figure}[htp]
  \includegraphics[width=\columnwidth]{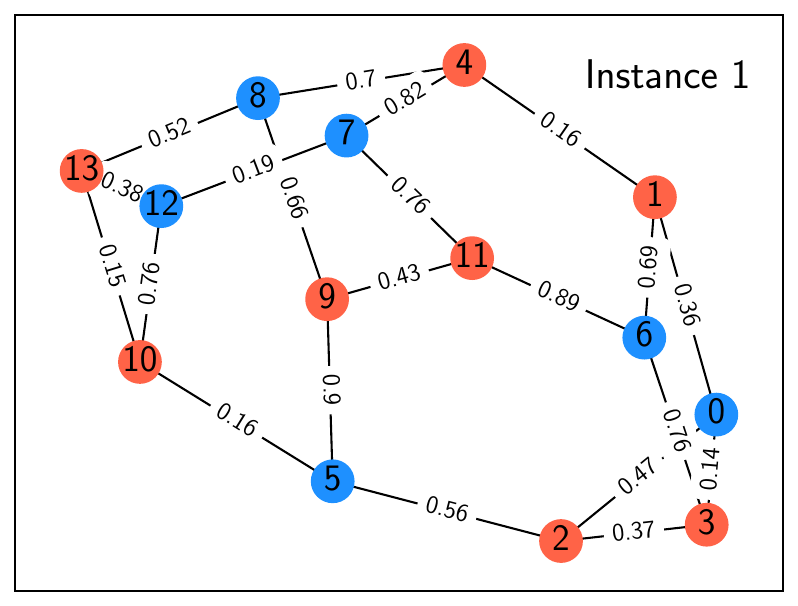}%
\vspace{-5mm}
\caption{
A specific MaxCut instance with $N=14$ vertices considered in our work. The red and the blue colors indicate the two optimal partitions that solve the MaxCut problem.
This instance has been examined in full detail also in Ref.~\cite{zhou_quantum_2020}, proving particularly difficult for a linear-schedule QA approach due to a very small minimum gap.}
\label{fig:graph_inst0}
\end{figure}

The variational energy for this digital optimal control problem is given by:
\begin{equation} \label{eqn:E_dQA}
E_{\Ptrot}^{\dQA}(\C) = \langle \Psi_{\Ptrot}^{\dQA}(\btheta)| \Ho_z | \Psi_{\Ptrot}^{\dQA}(\btheta)\rangle  \;,
\end{equation}
where the $2\Ptrot$ parameters $\btheta=(\btheta^x,\btheta^z)$ are regarded as functions of the $2\Nbasis+1$ coefficients $\C=(\rmC_0,\C^x,\C^z)$.
Besides the clear benefit of digitization, which is the lower computational cost of a time-evolution in a classical simulation, the digitized dynamics for a two-body $\Ho_z$ admits a standard decomposition into elementary one- and two-qubit gates, amenable to direct implementation on gate-based devices~\cite{Nature_dQA}.
As an important side benefit, we can calculate {\em analytically} the gradients of the variational energy with respect to the variational parameters $\C$ \cite{KHANEJA_2005, DEFOUQUIERES_2011},
see App.~\ref{app:gradient} for details,
hence allowing for faster gradient-based optimization routines. 
If $\C^*$ denotes the (local or global) optimal parameters found, the corresponding variational energy is given by:
\begin{equation}  \label{eqn:Efin_dQA}
E_{\Ptrot}^{\fin} =  E_{\Ptrot}^{\dQA}(\C^*)  \;.
\end{equation}

Our practical implementation of the F-CRAB algorithm relies on an iterative procedure that is particularly effective in obtaining smooth solutions.
In operational terms, these are characterized by a regular \emph{smooth} pattern of each of the two optimal vectors $(\theta^x_1\cdots\theta^x_{\Ptrot})$ and $(\theta^z_1\cdots\theta^z_{\Ptrot})$, when plotted vs the layer index $m=1\cdots\Ptrot$.
The rationale behind the implementation of an iterative scheme is that a direct optimization of $2 \Nbasis+1$ coefficients often gets trapped in a low-quality local minimum for large $\Nbasis$. Even when optimal or nearly-optimal solutions are found, these are characterized by irregular non-smooth behavior, namely exhibiting sharp variations for a small increase of $m$. 

To address this issue, we start by optimizing the $\Nbasis = 2$ case. 
By following the CRAB prescription, the $\Nbasis = 2$ frequencies are randomly generated for $n_r$ times, a new optimization of the $2\Nbasis +1 = 5$ coefficients is performed for each realization, and the best result in terms of variational energy is selected.
Then, we optimize the cases ${\mathrm{N}_c'} > \Nbasis$ according to the following procedure. We use the optimal $2\Nbasis +1$ parameters obtained in the previous iteration as a warm start for the optimization of the new $2{\mathrm{N}_c'}+1$ coefficients, initializing the remaining $2({\mathrm{N}_c'} - \Nbasis)$ coefficients to zero.
Crucially, at each iterative step, the old $\Nbasis$ frequencies are kept fixed, whereas the new ${\mathrm{N}_c'} - \Nbasis$ frequencies are once again randomly generated for $n_r$ times, selecting the optimal outcome with minimal variational energy.

We iterate this procedure, increasing the number of frequencies up to 
$\Nbasis = \Ptrot/2$. 
We choose this threshold as for $\Nbasis > \Ptrot/2$ we do not observe any significant performance improvement, compared with the increase of computational cost.
This iterative optimization involves a computational overhead; however, for large $\Ptrot$, it turns out to be sufficient to increase $\Nbasis$ in steps of $10$ in order to obtain smooth optimal curves. 

The occurrence of irregular solutions in a direct optimization for F-CRAB can be intuitively understood by considering that, for large $\Nbasis$, the Fourier expansion involves high-frequency modes that may induce fast oscillations in the optimal parameters $\btheta$. 
Nevertheless, limiting the expansion to low-frequency modes does not empirically guarantee a good performance in terms of variational energy. 
The relation between irregular and smooth solutions is analyzed in greater detail in Sec. \ref{sec:smoothness_and_annealing_path}.

Such an iterative procedure, as it turns out, is not needed for C-CRAB. Indeed, this method provides high-quality smooth curves without any need for a warm-start: for each value of $\Ptrot$ and of $\Nbasis$, we set a linear dQA initial condition for the optimization, 
i.e.\ $\rmC_0 = 1$ and $\C^x=\C^z=0$.  
This difference can be explained by noting that the Chebyshev polynomials of degree $n$ oscillate slower compared to the $n$-th Fourier modes. 

\subsection{QAOA} \label{sec:QAOA}
The previous approach to the optimal control problem leads to a form of the variational state which is identical to that of the
Quantum Approximate Optimization Algorithm (QAOA) by 
Farhi {\em et al.}~\cite{Farhi_arXiv2014}.
The QAOA state is expressed as:
\begin{equation} \label{eqn:QAOA_state}
|\Psi^{\QAOA}_{\Ptrot}(\btheta)\rangle =
\Uo(\theta^x_\Ptrot,\theta^z_\Ptrot) \cdots 
\Uo(\theta^x_1,\theta^z_1) \, |\Psi_0\rangle \;,
\end{equation}
where, again,
\begin{equation} \label{eq:U_m}
    \Uo(\theta^x_m,\theta^z_m) =  \nep^{-i \theta^x_m \Ho_x} \, \nep^{-i \theta^z_m \Ho_z} \;.
\end{equation}
The difference is that, here, $\btheta$ are interpreted as $2\Ptrot$ free variational parameters, without any explicit reference to time-dependent smooth control fields.  Although the form of the state in Eq.~\eqref{eqn:QAOA_state}
is identical to that of the digitized optimal control state in Eq.~\eqref{eqn:dQA_state}, the idea behind the two approaches is, in principle, different: 
the first strategy is based on the optimization of the expansion coefficients over a truncated basis of smooth driving fields,
while the QAOA strategy is in principle agnostic of any smoothness property of optimal solutions. 
Nonetheless, smooth optimal solutions have been observed in the QAOA framework in several case studies~\cite{zhou_quantum_2020, Wauters_PRR2020, Mele_PRA2022, LGT_QAOA_2021, Farhi_2022, mbeng_quantum_2019, Sack2021quantumannealing, Torta_PRB2023, Wurtz2022counterdiabaticity, Pagano_PNAS2020}.
Here, the variational energy is defined as:
\begin{equation} \label{eqn:E_QAOA}
E_{\Ptrot}^{\QAOA}(\btheta) = \langle \Psi_{\Ptrot}^{\QAOA}(\btheta)| \Ho_z | \Psi_{\Ptrot}^{\QAOA}(\btheta)\rangle  \;,
\end{equation}
and if $\btheta^*$ denotes a (local or global) optimal solution, then the variational energy is given by:
\begin{equation}  \label{eqn:Efin_QAOA}
E_{\Ptrot}^{\fin} =  E_{\Ptrot}^{\QAOA}(\btheta^*)  \;.
\end{equation}
The minimization of this cost function is, in general, a highly non-trivial task since random-start local optimization routines tend to get trapped into one of the many local minima of the $2\Ptrot$-dimensional search space~\cite{Bittel_PRL2021}.
Moreover, most of this landscape is plagued by the phenomenon of 
{\em barren plateaus}~\cite{McClean_NatCom2018},
whereby the gradients of the function to be minimized are exponentially small in the number of qubits.

Notice that digitized CRAB not only biases the search towards smooth solutions but also effectively explores a restricted portion of the QAOA energy landscape. Combined with an iterative procedure such as the one described for F-CRAB, digitized optimal control methods represent an effective way to optimize the QAOA variational parameters.
Previous research on QAOA proposed alternative iterative methods to cope with the difficulties of an unfavorable optimization landscape, which are briefly reviewed below and compared to CRAB in Section~\ref{subsec:comparison}.

\subsubsection{Interpolation schemes}

\textbf{INTERP.} 
One natural strategy is to proceed iteratively from a lower-$\Ptrot$ minimum to larger $\Ptrot'$, for instance $\Ptrot'=\Ptrot+1$, by appropriate interpolation of
the known solution. More in detail, suppose that $\btheta^{(*,\Ptrot)}$ is a previously determined (local or global) optimal schedule. 
Then, we can construct a candidate solution for $(\Ptrot+1)$ with the following procedure. 
Define an initial $2(\Ptrot+1)$-dimensional vector $\btheta^{(0,\Ptrot+1)}$ as follows 
(the same expression applies both to $\theta^x_m$ and $\theta^z_m$, hence we omit indicating the label x/z):
\begin{equation}
\theta^{(0,\Ptrot+1)}_m = \frac{m-1}{\Ptrot} \theta^{(*,\Ptrot)}_{m-1} + \frac{\Ptrot-m+1}{\Ptrot} \theta^{(*,\Ptrot)}_{m} \;, 
\end{equation}
where $m=1,\cdots,\Ptrot+1$. 
Notice that, although in principle referenced in the formula, $\theta^{(*,\Ptrot)}_{0}$ and $\theta^{(*,\Ptrot)}_{\Ptrot+1}$
are always multiplied by zero, hence they are irrelevant. 
In particular, $\theta^{(0,\Ptrot+1)}_1=\theta^{(*,\Ptrot)}_1$ and $\theta^{(0,\Ptrot+1)}_{\Ptrot+1}=\theta^{(*,\Ptrot)}_{\Ptrot}$.
The candidate solution $\btheta^{(0,\Ptrot+1)}$ is utilized as a warm start for a local gradient-based optimization, yielding the optimal $\btheta^{(*,\Ptrot+1)}$.
This strategy was introduced in Ref.~\cite{zhou_quantum_2020} and dubbed INTERP. In practice, it can be applied starting e.g.\ from $\Ptrot=2$,  and iterating the algorithm for increasing $\Ptrot=3,4,\cdots$ up to the desired final value of $\Ptrot_{\max}$. Alternatively, it can be slightly modified to allow for $\Ptrot' = \Ptrot + \Delta \Ptrot$, with $\Delta \Ptrot > 1$.

\textbf{LogINTERP.} 
An alternative construction, introduced in Ref.~\cite{mbeng_quantum_2019}, proceeds by {\em doubling} $\Ptrot$, specifically by using an optimal solution at depth $\Ptrot$ to seed the optimization for depth $2\Ptrot$. We apply it in two variants, each adapted to different boundary conditions for $\btheta^z$ and $\btheta^x$.
Whenever the parameters are likely bound to vanish at the initial time-step, as for $\btheta^z$, we take as an initial guess:
\begin{equation}  \label{eqn:LogINTERP_odd}
\left\{
\begin{array}{ll}
\theta^{(0,2\Ptrot)}_{2m} = \theta^{(*,\Ptrot)}_{m}  	
& \hspace{2mm} m=1\cdots \Ptrot \vspace{4mm} \\
\theta^{(0,2\Ptrot)}_{2m-1} = \frac{1}{2} ( \theta^{(*,\Ptrot)}_{m-1} + \theta^{(*,\Ptrot)}_{m} ) 
& \hspace{2mm} m=1 \cdots \Ptrot 
\end{array}
\right. \;, 
\end{equation}
where the boundary condition $\theta^{(*,\Ptrot)}_0=0$ should be used.
Alternatively, when the vanishing boundary condition is at the final time-step, as for $\btheta^x$, we take as initial guess:
\begin{equation} \label{eqn:LogINTERP_even}
\left\{
\begin{array}{ll}
\theta^{(0,2\Ptrot)}_{2m-1} = \theta^{(*,\Ptrot)}_{m}  	
& \hspace{2mm} m=1\cdots \Ptrot \vspace{4mm} \\
\theta^{(0,2\Ptrot)}_{2m} = \frac{1}{2} ( \theta^{(*,\Ptrot)}_{m} + \theta^{(*,\Ptrot)}_{m+1} ) 
& \hspace{2mm} m=1\cdots \Ptrot
\end{array}
\right. \;,
\end{equation}
where the boundary condition $\theta^{(*,\Ptrot)}_{\Ptrot+1}=0$ should be used. These qualitative features in the boundary conditions for the optimal angles have been observed in QAOA-like numerics for a breadth of different tasks~\cite{Mbeng_PhDThesis2019New,Pagano_2020,zhou_quantum_2020, farhi2020quantum,Wurtz_2022,Crooks_arXiv2018,QAOApatternbrady2021behavior}.
The optimal $\btheta^{(*,2\Ptrot)}$ is found by applying a local gradient-based optimization starting from the guess $\btheta^{(0,2\Ptrot)}$.

\subsubsection{Fourier-based schemes}

\textbf{FOURIER.}
A strategy based on expansion in Fourier modes was introduced in Ref.~\cite{zhou_quantum_2020}, denoted as FOURIER.
It is based on the following parameterization:
\begin{equation}
\left\{
\begin{array}{l}
    \theta^z_m = \displaystyle \sum_{n=1}^{\Nbasis} \rmC^z_n \textstyle \sin\big((n-\frac{1}{2})(m-\frac{1}{2})\frac{\pi}{\Ptrot}\big) \vspace{3mm} \\
    \theta^x_m = \displaystyle \sum_{n=1}^{\Nbasis} \rmC^x_n \textstyle \cos\big((n-\frac{1}{2})(m-\frac{1}{2})\frac{\pi}{\Ptrot}\big)
\end{array}
\right. \;.
\end{equation}
Observe the similarities with the F-CRAB approach we discussed previously. 
Note, however, the absence of a linear dQA-like term multiplying the Fourier decomposition,
and also the fact that the Fourier frequencies are fixed, and randomness is only used in searching for the optimal coefficients. 
Different variants of this strategy postulate how the final required $\Ptrot_{\max}$ is reached starting from lower values, 
and how many Fourier modes are allowed, either $\Nbasis=\Ptrot$, or a fixed predetermined value of $\Nbasis<\Ptrot$. 
Details are given in the original Ref.~\cite{zhou_quantum_2020}, and summarized in App.~\ref{app:FOURIER} for convenience.

\begin{figure}{}
\centering
\includegraphics[width=1\columnwidth]{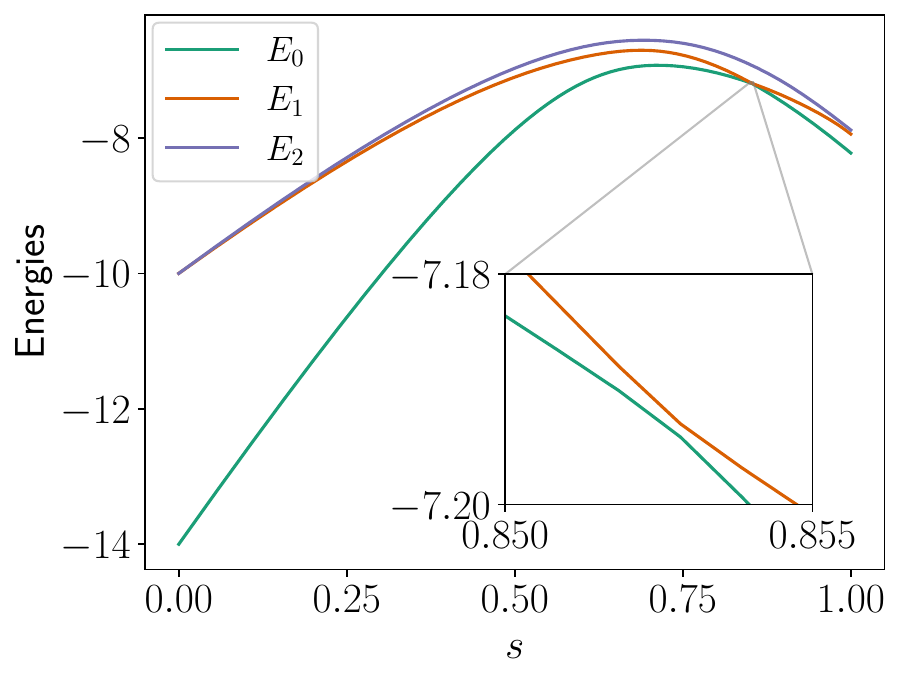}
\caption{Low-energy spectrum of the annealing Hamiltonian $s\Ho_z+(1-s)\Ho_x$ as a function of $s=t/\tau$, for the MaxCut instance $1$.
In the inset, we zoom on the closing gap, which is of order $10^{-3}$.}
\label{fig:spectrum_ann}
\end{figure}

\begin{figure}[htp]
\includegraphics[width=\columnwidth]{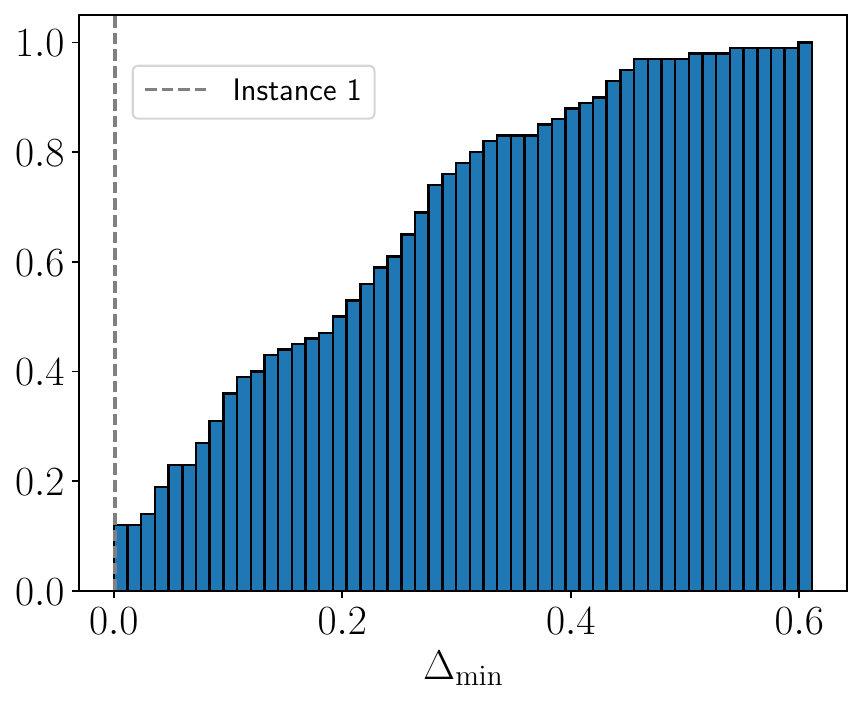}
\vspace{-5mm}
\caption{Cumulative distribution of the minimum gaps of the annealing Hamiltonian for 100 weighted $3-$regular graphs of $N=14$ vertices. A vertical dashed line indicates the small minimum gap for instance 1: similarly hard instances are selected for our numerics. To compute the gap, we considered a linear annealing schedule.
}
\label{fig:graph_gaps_instances}
\end{figure}

\begin{figure}[htp]
\includegraphics[width=\columnwidth]{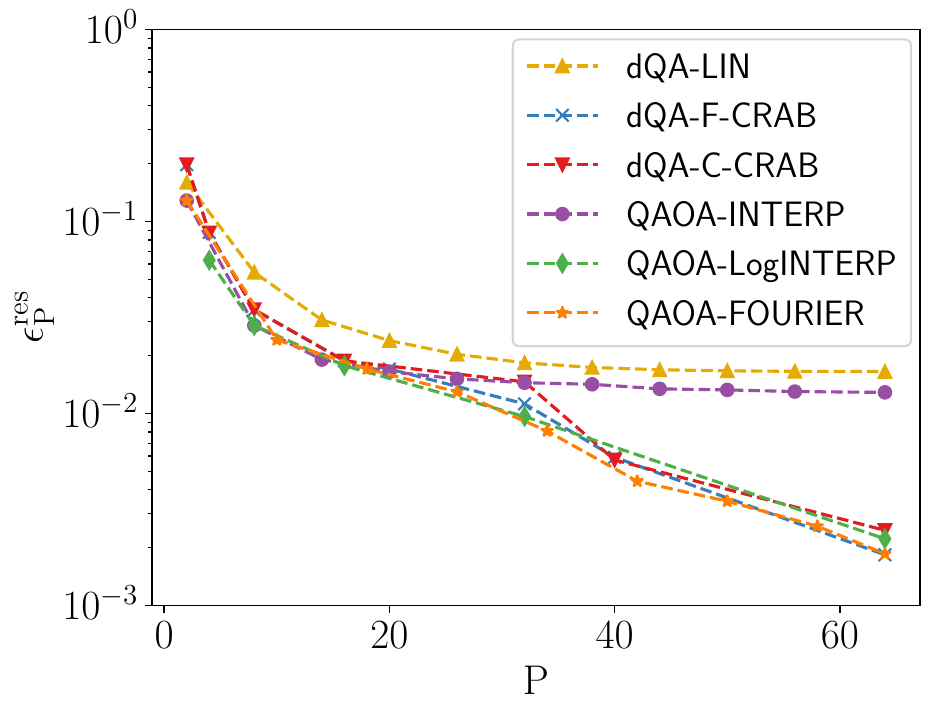}\\
\includegraphics[width=\columnwidth]{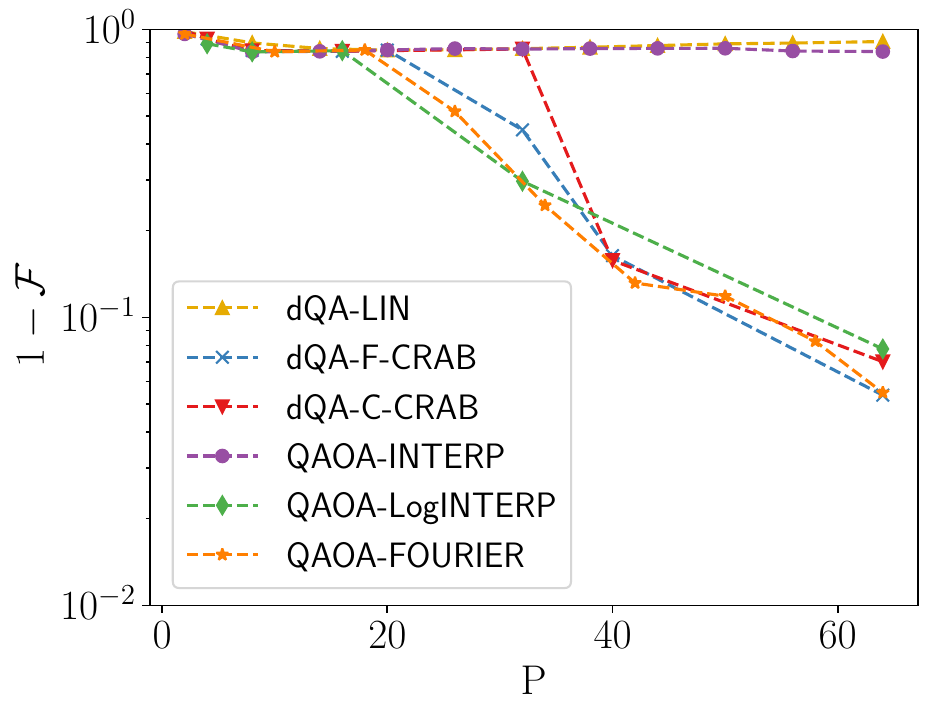}
\caption{Residual energies (top) and infidelity (bottom) as a function of $\Ptrot$ for the instance $1$ and for different methods. Lines are guide to the eye. Infidelity and residual energy behave consistently. 
We see that digitized linear-schedule Quantum Annealing (dQA-LIN), and QAOA-INTERP get stuck in a local minimum for large $\Ptrot$. 
The four methods that perform better are dQA-F-CRAB, dQA-C-CRAB, QAOA-FOURIER and QAOA-LogINTERP.
}
\label{fig:fidelity_vs_eres}
\end{figure}

\section{The MaxCut problem} \label{sec:maxcut}

The MaxCut problem is a classical combinatorial problem in the NP-complete complexity class. 
It is defined on a graph $\mathcal{G}$ of $N$ vertices, with edges denoted as $\mathcal{E}$. We consider weighted graphs, i.e., each edge $(j j')\in \mathcal{E}$ is associated to a weight $J_{jj'}$ randomly drawn in the interval $[0,1]$ with uniform distribution. 
The goal is to find a path which cuts the largest possible number of edges (more precisely, the largest sum of weights $J_{jj'}$) 
so that the graph $\mathcal{G}$ is partitioned into two subgraphs. One way to proceed is to assign to each vertex a spin $\uparrow$ or $\downarrow$, and cut the edges that link two vertices with different spin. 
In this way, we can reformulate the problem in terms of the ground state search of an antiferromagnetic spin Hamiltonian
\begin{equation}
\label{eqn:H_maxcut}
\Ho_{\target} = \Ho_z = \frac{1}{2} \sum_{(j j')\in \mathcal{E}} J_{jj'} (\PauliSigma^z_j \PauliSigma^z_{j'} - 1) \;,
\end{equation}
where each term in the sum has been rescaled and shifted to let the ground state energy coincide with (minus) the maximum number of cut edges 
(for the unweighted case with $J_{jj'}=1$).
The maximum energy is $E_{\max} = 0$, associated with the configuration where all the spins are identical (ferromagnetic state).

Figure \ref{fig:graph_inst0} illustrates a specific MaxCut instance, previously analyzed in  Ref.~\cite{zhou_quantum_2020}. 
This instance is particularly hard for a linear-schedule QA approach, as one quickly discovers by plotting the instantaneous spectrum of the annealing Hamiltonian 
$\Ho(s) = s\Ho_z+(1-s)\Ho_x$ with $s=t/\tau$, as shown in Fig.~\ref{fig:spectrum_ann}. In our work, we focus on a set of similarly hard instances of the MaxCut problem with $N=14$ vertices.
To select hard instances, we first generate several random $3-$regular weighted graphs and compute the instantaneous spectrum 
of the linear annealing Hamiltonian.
In Fig.~\ref{fig:graph_gaps_instances} we show the normalized cumulative distribution of the minimum gaps for $100$ random instances. 
We define as hard the instances having a minimum gap $\Delta_{\min}$ of the order $\sim 10^{-3}$. 

Since linear QA is not expected to solve these instances without resorting to very large values of the annealing time $\tau$, a MaxCut optimal solution may be obtained more effectively by employing the quantum algorithms introduced in Section~\ref{sec:methods}. 
We tested our methods on a total of $10$ hard instances.

\begin{figure*}[htp]
\captionsetup[subfigure]{labelformat=empty}
\subfloat[]{%
  \includegraphics[width=\columnwidth]{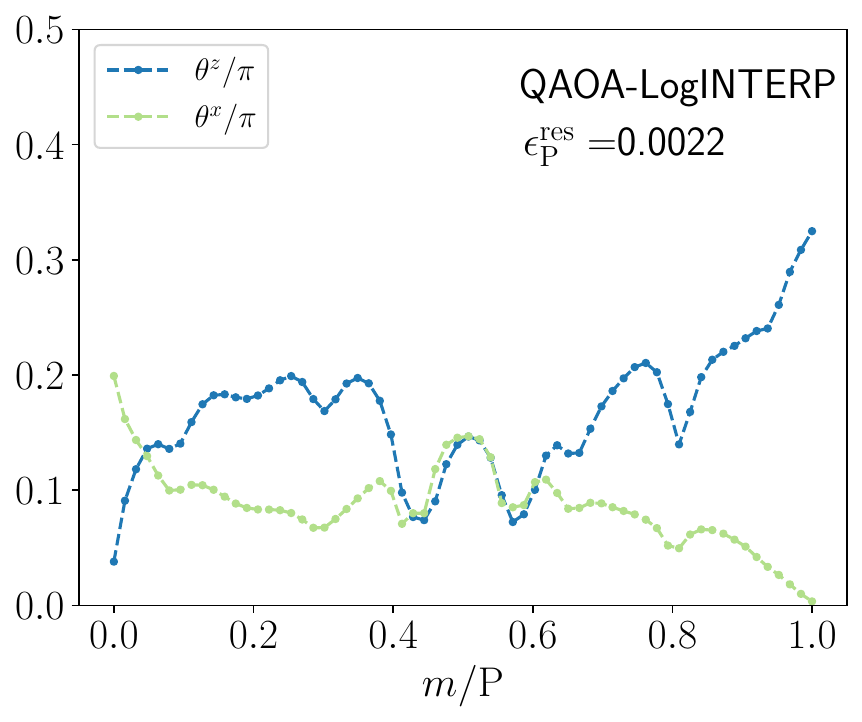}%
  \includegraphics[width=\columnwidth]{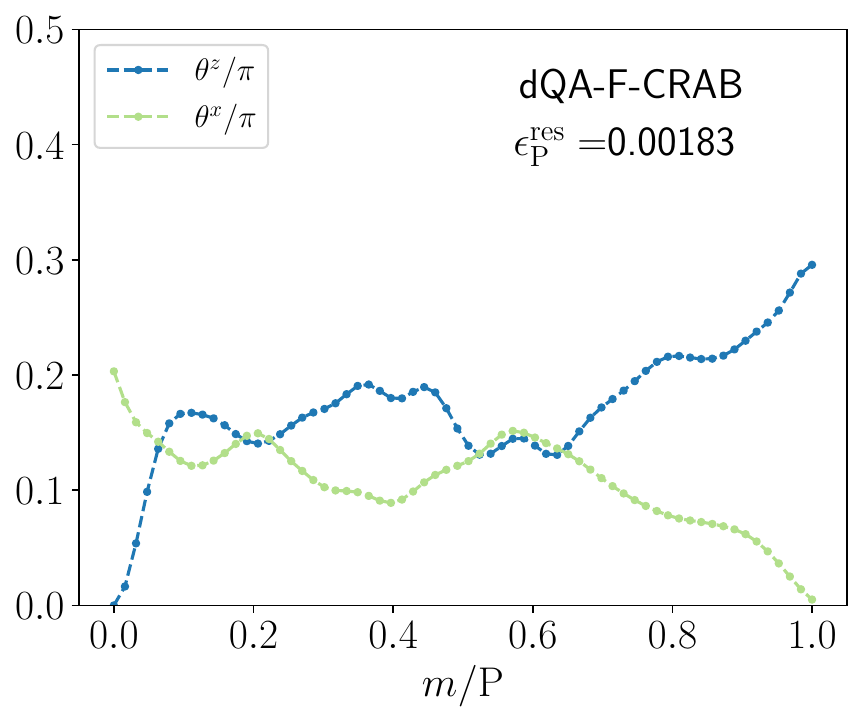}%

}
\vspace{-9mm}
\subfloat[]{%

  \includegraphics[width=\columnwidth]{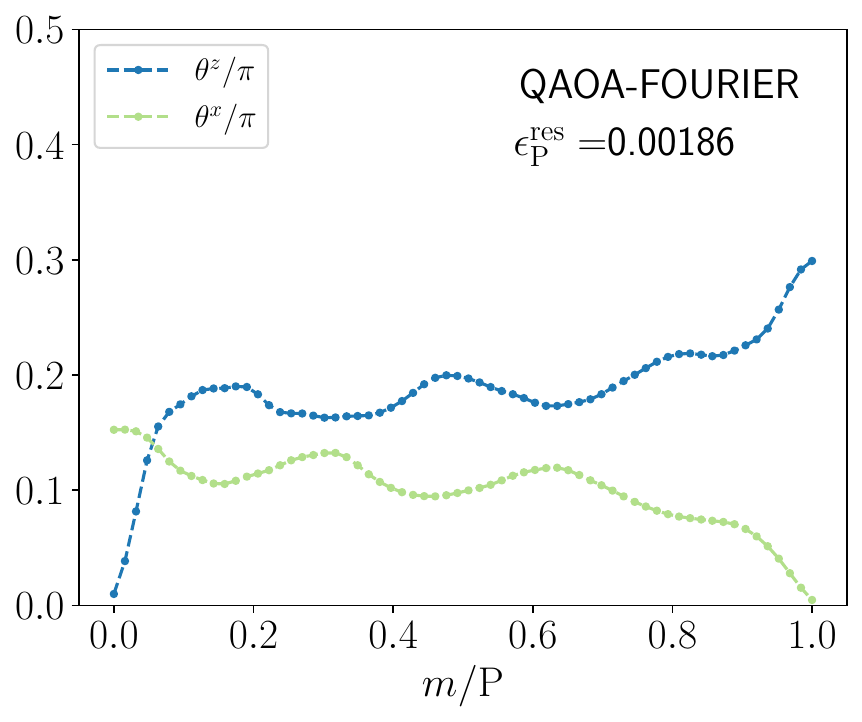}%
  
  \includegraphics[width=\columnwidth]{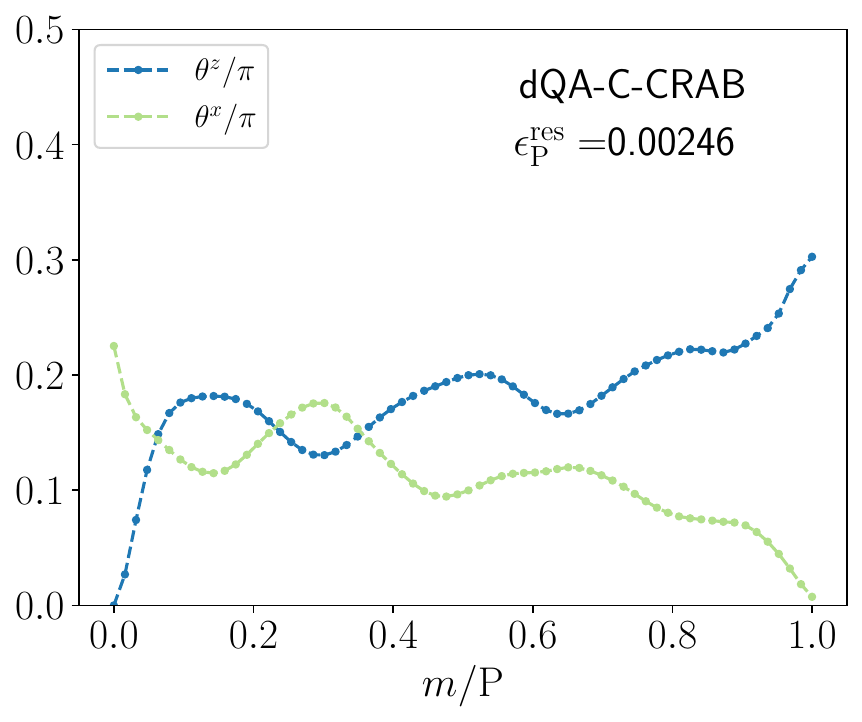}%

}
\vspace{-7mm}
\caption{Optimal parameters obtained with different methods for instance $1$ and $\Ptrot=64$. For C-CRAB, F-CRAB, and FOURIER, the number of functional coefficients is $\Nbasis = \Ptrot/2 = 32$.
} 
\label{fig:parameters_zhou}
\end{figure*}

As a figure of merit for the effectiveness of our techniques, we will mostly use the residual energy, defined as a rescaled version of the variational energy in 
Eqs.~\ref{eqn:Efin_dQA},~\ref{eqn:Efin_QAOA}:
\begin{equation}
\label{eqn:res_energy}
\epsilon^{\res}_{\Ptrot} = \frac{E_{\Ptrot}^{\fin} - E_{\min}}{E_{\max} - E_{\min}} \;,
\end{equation}
where $E_{\min}$ and $E_{\max}$ are the minimum (ground-state) and maximum energy of the problem Hamiltonian in Eq.~\ref{eqn:H_maxcut}. 
A traditional figure of merit used in the MaxCut literature~\cite{zhou_quantum_2020} is $r=E_{\Ptrot}^{\fin}/E_{\min}$. 
Notice that $E_{\min}<0$ and $0<r<1$, with $r$ approaching $1$ as the solution improves.
The two quantities are related as $1-r =\epsilon^{\res}$ since $E_{\max}=0$ when $J_{ij}>0$ (ferromagnetic state). 

An alternative figure of merit is the fidelity \cite{Nielsen_Chuang:book}
between the final (optimal) state $|\Psi_{\Ptrot} \rangle$, and the exact ground state of $\Ho_{\target}$:
\begin{equation}
\mathcal{F} = |\langle \Psi_{\mathrm{gs}} | \Psi_{\Ptrot} \rangle |^2 \;.
\end{equation}
In principle, very good solutions in terms of residual energy 
$\epsilon^{\res}_{\Ptrot}$ might have low fidelity, if the lowest excited states of the target Hamiltonian are very close in energy to the ground state. 
In the context of the MaxCut problems under consideration, we note that the infidelity, expressed as $1-\mathcal{F}$, gives results that are qualitatively consistent with the residual energy $\epsilon^{\res}_{\Ptrot}$ for all the methods investigated. 

\section{Results}\label{sec:results}
\subsection{Comparison of different methods}
\label{subsec:comparison}

In this section, we compare the performance of the different methods illustrated in Sec.~\ref{sec:methods}. These include dQA with a linear schedule (dQA-LIN) or using 
Fourier- or Chebyshev-based CRAB (dQA-F-CRAB, dQA-C-CRAB) 
and QAOA-based techniques, in particular QAOA-INTERP, QAOA-LogINTERP, and QAOA-FOURIER. 
For ease of presentation, we show results focusing on the instance $1$ depicted in Fig.\ref{fig:graph_inst0}, the main one considered also in Ref.~\cite{zhou_quantum_2020}.
We verified that our results are valid for all the other hard instances analyzed (See Appendix \ref{app:other_instances} for further analysis).

In Fig.\ref{fig:fidelity_vs_eres}, we compare residual energy and infidelity values obtained using the different methods.
One immediately observes that, for shallow circuits, $\Ptrot\le 20$, all methods behave roughly in the same way. 
For larger values of $\Ptrot$, two methods get trapped into some kind of plateau: dQA-LIN and QAOA-INTERP. 
Indeed, due to the small value of the minimum gap of the selected instance, the performance of a linear-schedule digitized QA (dQA-LIN) is rather poor. 
Perhaps surprisingly, a similar saturation occurs for QAOA-INTERP~\cite{zhou_quantum_2020}.
On the contrary, the two CRAB-based dQA, as well as QAOA-LogINTERP and QAOA-FOURIER perform much better, 
as witnessed by decreasing values of
$\epsilon^{\res}_{\Ptrot}$ and infidelity $1-\mathcal{F}$, 
for increasingly larger $\Ptrot$.
These methods show similar performance, and all provide smooth optimal schedules, as shown in Fig.\ref{fig:parameters_zhou}. 

\subsection{Eigenstate population and shortcut to adiabaticity}
Here, we analyze the details of the dynamics implemented by the different methods presented above. 
In order to describe the time evolution of the system, we introduce an effective Hamiltonian that generates the digital dynamics:
\begin{equation} \label{eq:effective_ham}
    \Uo(\theta^x_m,\theta^z_m) =  \nep^{-i \theta^x_m \Ho_x} \, \nep^{-i \theta^z_m \Ho_z} \equiv 
    \nep^{-i \Ho^{\eff}_m} \;.
\end{equation}
The effective Hamiltonian $\Ho^{\eff}_m$ can be expressed in terms of $\Ho_x$, $\Ho_z$ and their nested commutators, using the Baker–Campbell–Hausdorff (BCH) formula: 
\begin{align} \label{eq:BCH_formula}
\Ho^{\eff}_m &= \theta^x_m \Ho_x + \theta^z_m \Ho_z - \frac{i}{2} \theta^x_m \theta^z_m [\Ho_x , \Ho_z] \nonumber \\
&\phantom{=} \hspace{2mm} +\frac{1}{12} (\theta^x_m)^2 \theta^z_m
[\Ho_x,[\Ho_x,\Ho_z]] \nonumber \\
&\phantom{=} \hspace{2mm} -\frac{1}{12} \theta^x_m (\theta^z_m)^2
[\Ho_z,[\Ho_x,\Ho_z]]
+ \cdots \;.
\end{align}

We will truncate the previous equation to the third order --- i.e., to the terms shown --- and perform exact diagonalization to obtain, at each time step $m$, the instantaneous eigenvectors of $\Ho^{\eff}_m$. 
The choice of the truncation order is justified by the fact that including higher order in Eq.~\eqref{eq:BCH_formula} does not significantly affect the values of the observables we computed. 

Clearly, each protocol is characterized by a different effective Hamiltonian.
The population of the $j$th instantaneous effective eigenstate $\ket{\phi^j_m}$ is defined as:
\begin{equation}
   p_j(m) = |\langle \phi^j_m | \Psi_{m} \rangle |^2 , 
   \label{eqn:populations}
\end{equation}
where, for each method, the state of the system at time step $m$ reads:
\begin{equation} 
|\Psi_{m}\rangle =
\Uo(\theta^x_ m,\theta^z_m) 
\cdots  
\Uo(\theta^x_1,\theta^z_1) \, |\Psi_0\rangle \;.
\end{equation}

\begin{figure}[htp]
\includegraphics[width=\columnwidth]{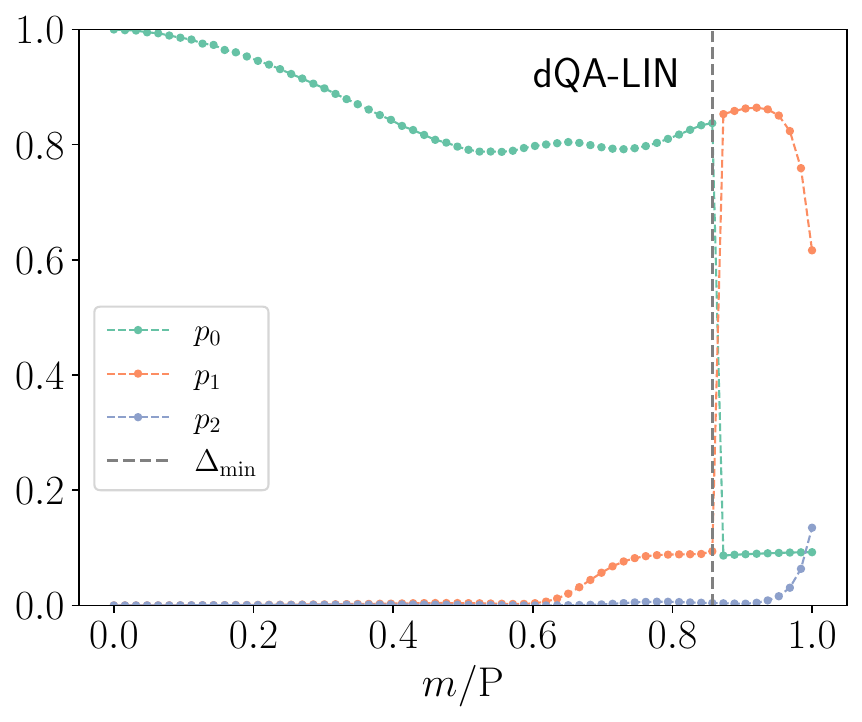}
\caption{ 
Population of the instantaneous ground state and of the first two excited states as a function of the index $m$ for the instance $1$ as obtained using dQA-LIN.
The vertical dashed line indicates where $\Ho^{\eff}_m$ attains its minimum spectral gap. 
The optimal time step for dQA-LIN is $\Delta_t^* = 0.78$.
The adiabatic protocol implemented by dQA-LIN fails to largely populate the ground state after the minimum gap. 
}
\label{fig:populations_dQA}
\end{figure}

\begin{figure*}[htp]
\captionsetup[subfigure]{labelformat=empty}
\subfloat[]{%
  \includegraphics[width=\columnwidth]{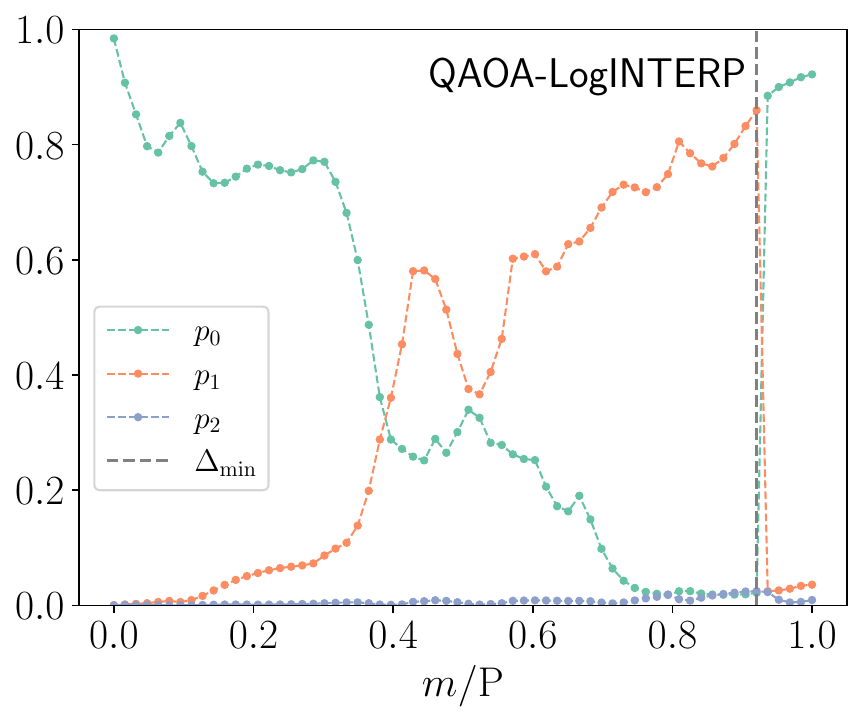}%
    \includegraphics[width=\columnwidth]{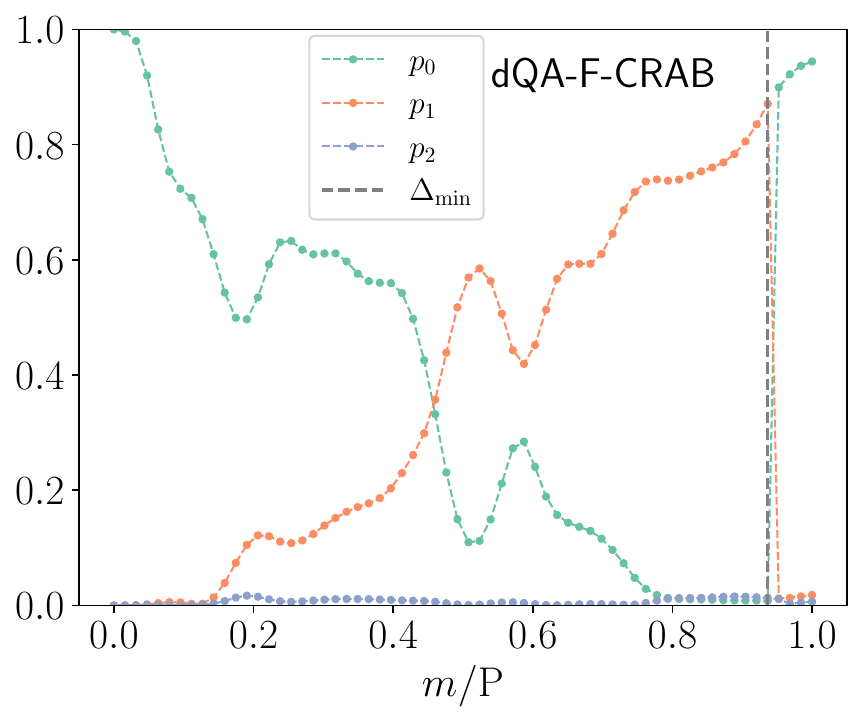}%
}
\vspace{-7mm}
\subfloat[]{%
\includegraphics[width=\columnwidth]{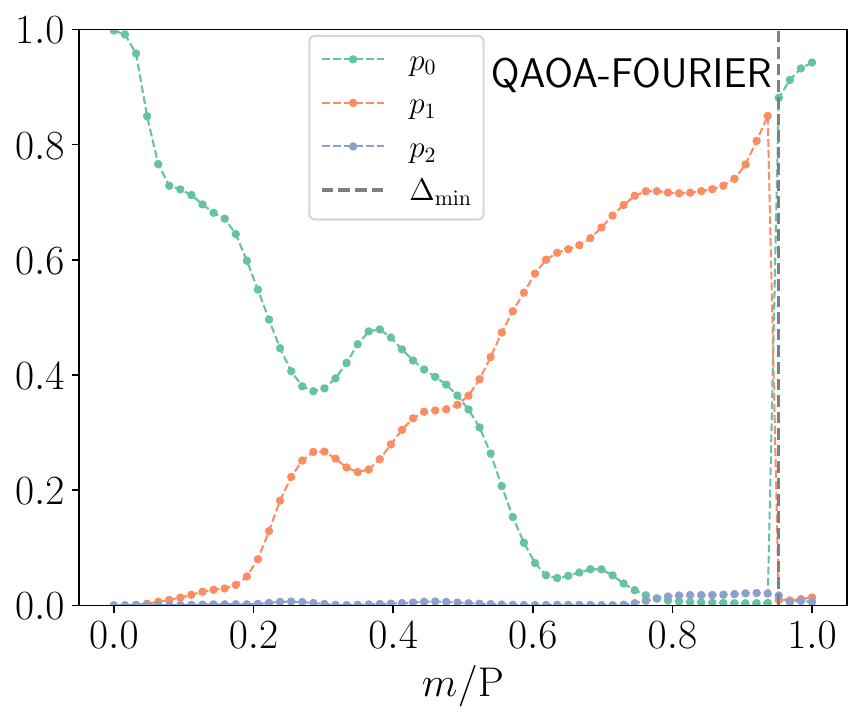}%
\includegraphics[width=\columnwidth]{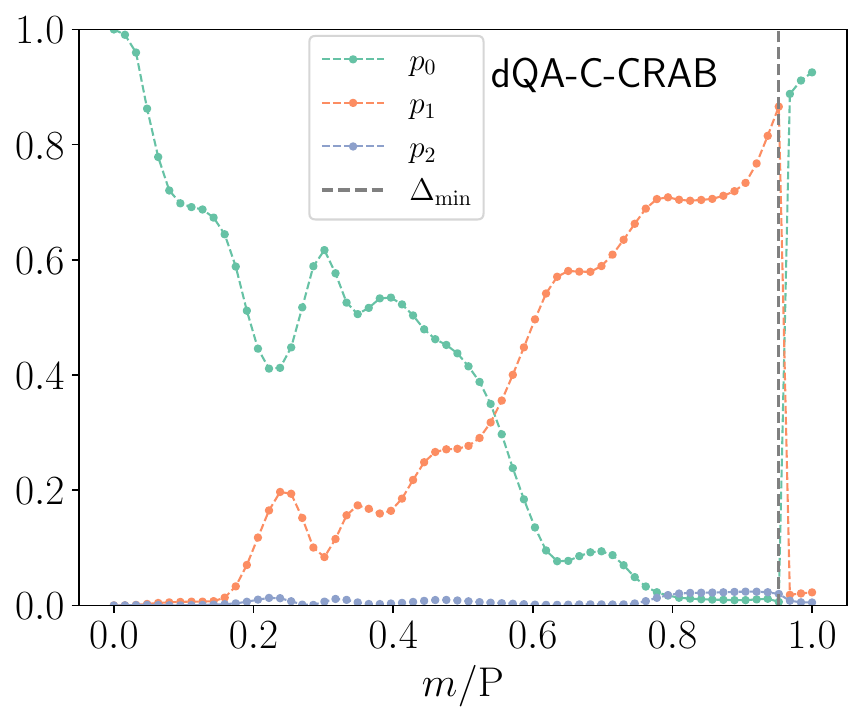}%
}
\vspace{-7mm}
\caption{Population of the instantaneous ground state and of the first two excited states as a function of the index $m$ for the instance $1$. 
We consider the same data as in Fig.~\ref{fig:parameters_zhou}. 
The vertical dashed lines indicate where the corresponding $\Ho^{\eff}_m$ attain their minimum spectral gap. 
In the four panels, each solution implements a different nonadiabatic protocol to reach the target ground state at $m=\Ptrot$. 
}
\label{fig:populations_P64}
\end{figure*}

In Fig.~\ref{fig:populations_dQA}, we focus on dQA-LIN, and we show the first three instantaneous eigenstate populations $p_j(m)$ for $j = 0,1,2$. 
Evidently, the system attempts to follow an adiabatic dynamics, 
maintaining a very high population $p_0(m)$ for the instantaneous ground state until a small instantaneous gap is encountered: here, the system undergoes a population inversion between the instantaneous ground state and the first excited state, which turns out to be the most populated state at the final 
$m = \Ptrot$. 
The four panels of Fig.~\ref{fig:populations_P64} show the same quantities referring to the four methods that provide better solutions for this instance. 
In these cases, before the system reaches the minimum gap, there is a smooth population inversion between the instantaneous ground state and the first excited state. 
Afterward, a second population inversion occurs in correspondence with the minimum spectral gap of $\Ho^{\eff}_m$, yielding a large final overlap between the state of the system and the target state. 
Remarkably, the system follows an STA path in order to achieve a large ground state population at the final time step $m = \Ptrot$.
These findings are consistent with the results of \cite{zhou_quantum_2020}, obtained using the first-order BCH formula and the QAOA-FOURIER method.

We verified that these four methods implement similar but \emph{quantitatively different} non-adiabatic protocols to solve the MaxCut problem. 
Due to the large dimension of the parameter space for $\Ptrot = 64$, one cannot exclude the existence of a path linking the minima in 
Fig.~\ref{fig:parameters_zhou} without encountering a barrier in the residual energy landscape.
However, in App.~\ref{app:Hessian}, we show that there is at least one direction along which the minima are not pairwise connected. 
This emphasizes the absence of a single broad minimum basin in the variational energy landscape to which all these optimal parameters belong.

We highlight that a similar 
non-adiabatic mechanism is observed for all the hard instances we examined.
Specifically, the optimal parameters derived from methods outperforming dQA-LIN consistently exhibit an STA mechanism to avoid the lack of performance due to population inversion at the minimum gap.
Interestingly, the gap of the effective Hamiltonian remains small regardless of the optimized digital dynamics. 
Rather than seeking an effective Hamiltonian with a large gap, the solution 
lies in inducing smooth population inversion to achieve a large final overlap with the target state. 
This trend holds across all the hard instances we considered. 

\subsection{Transferability among hard instances}
\begin{figure*}[htp]
\captionsetup[subfigure]{labelformat=empty}
\includegraphics[width=0.3333\textwidth]{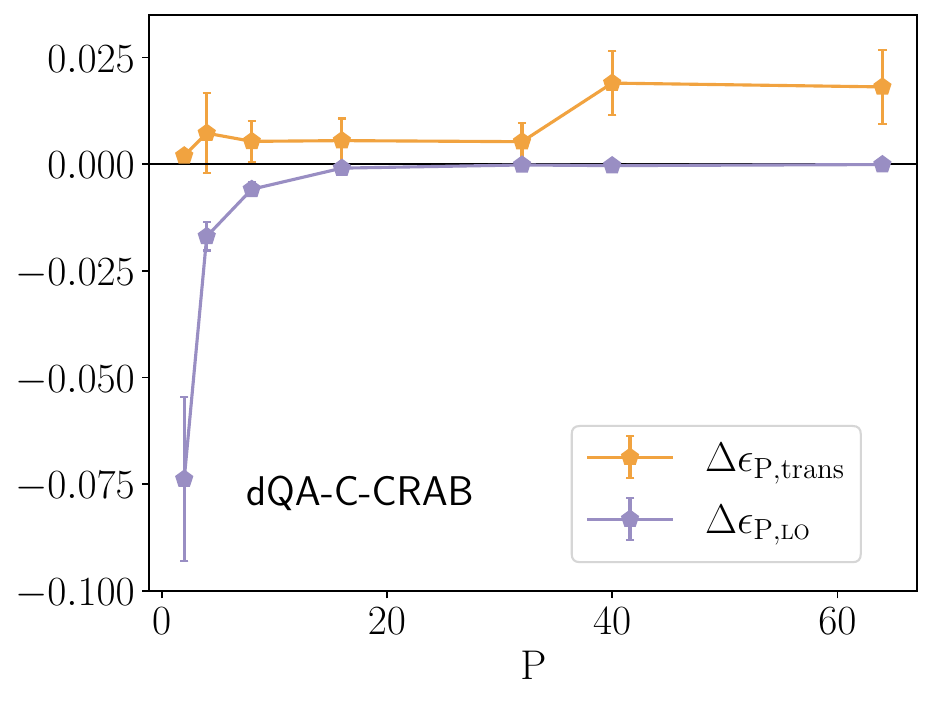}%
\includegraphics[width=0.3333\textwidth]{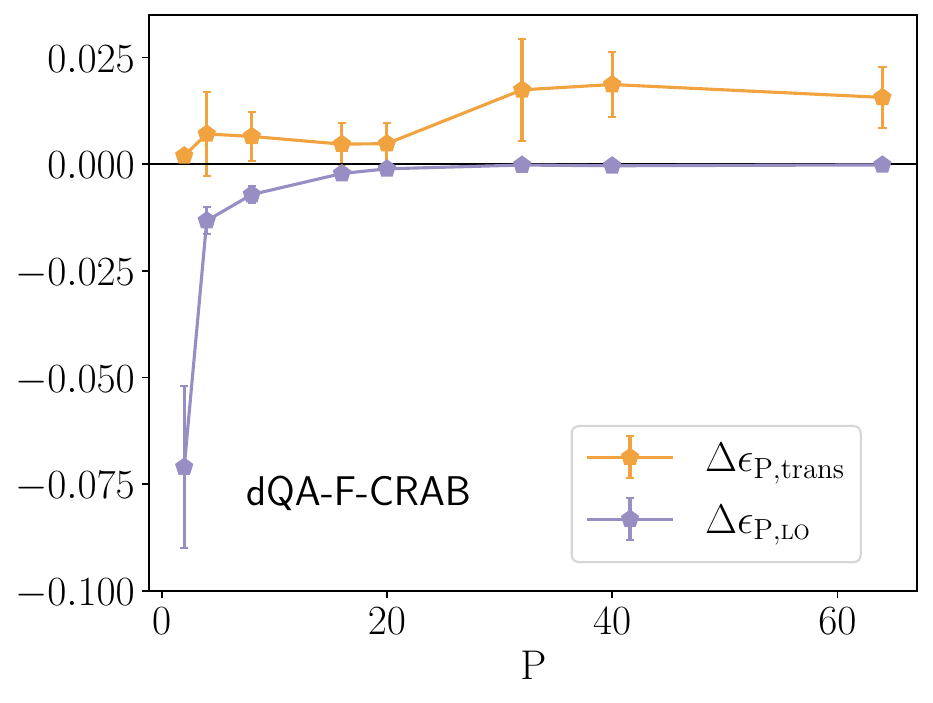}%
\includegraphics[width=0.3333\textwidth]{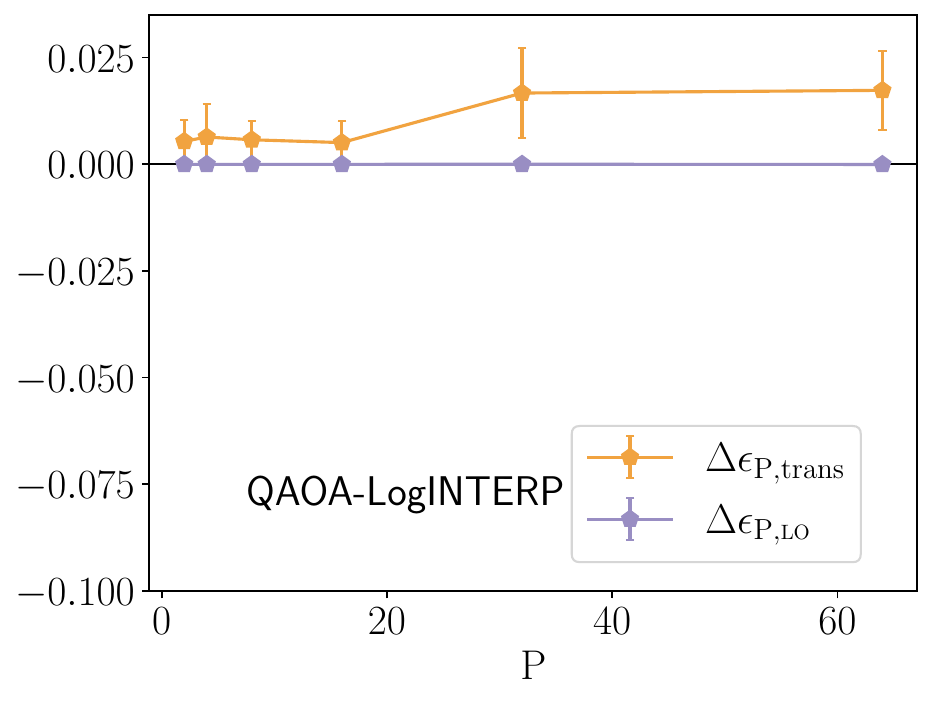}%
\caption{Plot of the mean differences 
$\Delta \epsilon_{\Ptrot, \text{trans}}$ and 
$\Delta \epsilon_{\Ptrot,\LO}$ as a function of $\Ptrot$ for three different methods, see the main text for details. 
We averaged over the $9$ hard instances over which the parameters of instance 1 have been transferred.  
The error bars indicate the standard deviation of the mean.
} 
\label{fig:transferability}
\end{figure*}

In this section, we focus on the transferability of the optimal parameters for dQA-C-CRAB, dQA-F-CRAB, and QAOA-LogINTERP, which appear to be the most effective methods (QAOA-FOURIER, as it turns out, is more computationally expensive).

Numerous studies have explored parameter concentration and transferability for both unweighted \cite{brandao2018fixed, Streif_2020} and weighted MaxCut \cite{Sureshbabu2024parametersettingin,10.1145/3584706}, often focusing on the regime of small $\Ptrot$ and large $N$. Special attention is devoted to typical instances drawn from a given distribution, also addressing problems beyond MaxCut or two-bodies interactions~\cite{QAOA_perceptron, Farhi2022quantumapproximate,9996946,basso_et_al:LIPIcs.TQC.2022.7}.

Here, we narrow our focus on hard instances of fixed and moderate size, exploring different circuit depths.
To probe the transferability of the optimal-control solutions, we proceed as follows. First, we evaluate, for fixed $\Ptrot$ and for each method, the residual energy of $10$ hard instances 
\begin{equation}
    E_{\Ptrot}^{\fin,k} = \langle \Psi_{\Ptrot}^{(k)}(\btheta^{(k)})| \Ho_z^{(k)} | \Psi_{\Ptrot}^{(k)}(\btheta^{(k)})\rangle  \;,
\end{equation}
where $k=1,\cdots,10$ labels the instances.

Next, we take the optimal parameters $\btheta^{(1)}$ of instance $k=1$,
and we compute the variational energy of the instances $k=2,\cdots,10$: 
\begin{equation}
    E_{\Ptrot,\text{trans}}^{\fin, k} = \langle \Psi_{\Ptrot}^{(k)}(\btheta^{(1)})| \Ho_z^{(k)} | \Psi_{\Ptrot}^{(k)}(\btheta^{(1)})\rangle  \;,
\end{equation}
where $|\Psi_{\Ptrot}^{(k)}(\btheta^{(1)})\rangle$ is the variational state for the instance $k$ with the optimal parameters of instance $k=1$, and $\Ho_z^{(k)}$ is the corresponding target Hamiltonian. 

Finally, we refine these transferred solutions by local optimization (LO), by considering the parameters $\btheta^{(1)}$ as a warm-start for a local optimization of the other instances $k=2,\cdots,10$, leading to final  
optimal parameters $\btheta_{\LO}^{(k)}$. 
With these transferred-and-optimized parameters, we compute the energy:
\begin{equation}
    E_{\Ptrot,\LO}^{\fin,k} = \langle \Psi_{\Ptrot}^{(k)}(\btheta_{\LO}^{(k)})| \Ho_z^{(k)} | \Psi_{\Ptrot}^{(k)}(\btheta_{\LO}^{(k)})\rangle  \;.
\end{equation}
Equation \eqref{eqn:res_energy} is then used to calculate the residual energies $\epsilon^{\res,k}_{\Ptrot}$, $\epsilon^{\res,k}_{\Ptrot,\text{trans}}$ and $\epsilon^{\res,k}_{\Ptrot,\LO}$. 

We summarize these results in Fig. (\ref{fig:transferability}), where we show the average differences
\begin{equation}
\begin{array}{lcl}
\Delta \epsilon_{\Ptrot, \text{trans}} &=&  
[\epsilon^{\res,k}_{\Ptrot,\text{trans}} - 
 \epsilon^{\res,k}_{\Ptrot}]_{\mathrm{av}} \vspace{3mm} \\
\Delta \epsilon_{\Ptrot,\LO} &=& 
[\epsilon^{\res,k}_{\Ptrot,\LO} - 
 \epsilon^{\res,k}_{\Ptrot}]_{\mathrm{av}} 
\end{array}
\end{equation}
as a function of $\Ptrot$ for dQA-C-CRAB, dQA-F-CRAB, and QAOA-LogINTERP.
The average is computed over the different hard instances we considered.
We observe that, for small $\Ptrot$, a direct transfer of the optimal parameters of instance $1$ provides residual energies very close to the optimal ones. 
On the other hand, at large $\Ptrot$ we see that a direct transfer of the optimal parameters of instance $1$ does not guarantee a good residual energy $\epsilon^{\res,k}_{\Ptrot,\text{trans}}$. 
At the same time, the transferred parameters turn out to be a good starting point, as witnessed by the fact that 
$\epsilon^{\res,k}_{\Ptrot,\LO}$ is, on average, 
comparable (or even better, for small $\Ptrot$) to  $\epsilon^{\res,k}_{\Ptrot}$, with the extra advantage of being computationally cheaper. 

\subsection{Smoothness of the solutions and continuous-time annealing}
\label{sec:smoothness_and_annealing_path}
\begin{figure}[htp]
\includegraphics[width=\columnwidth]{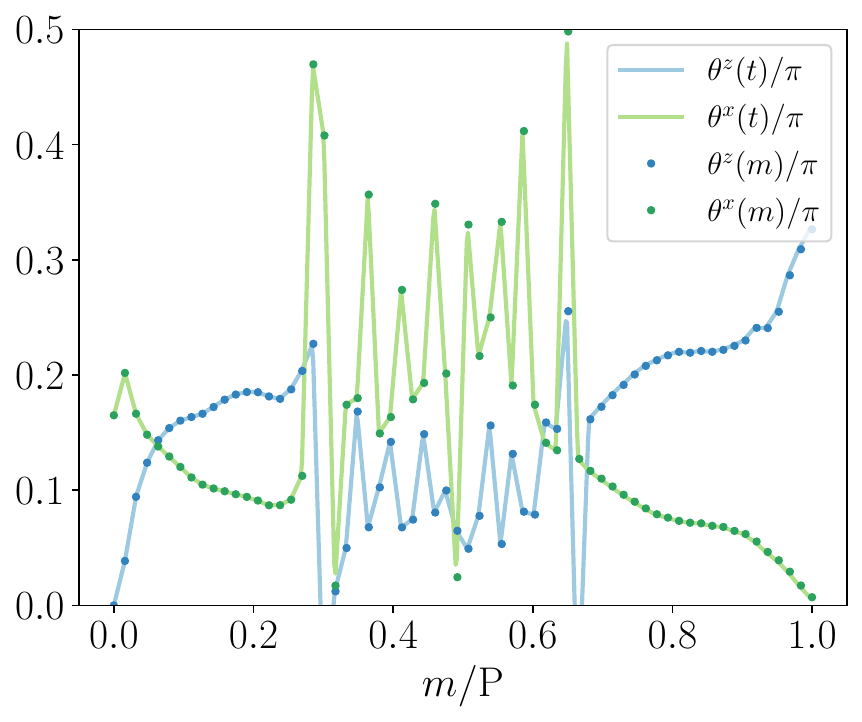} \\
\includegraphics[width=0.9\columnwidth]{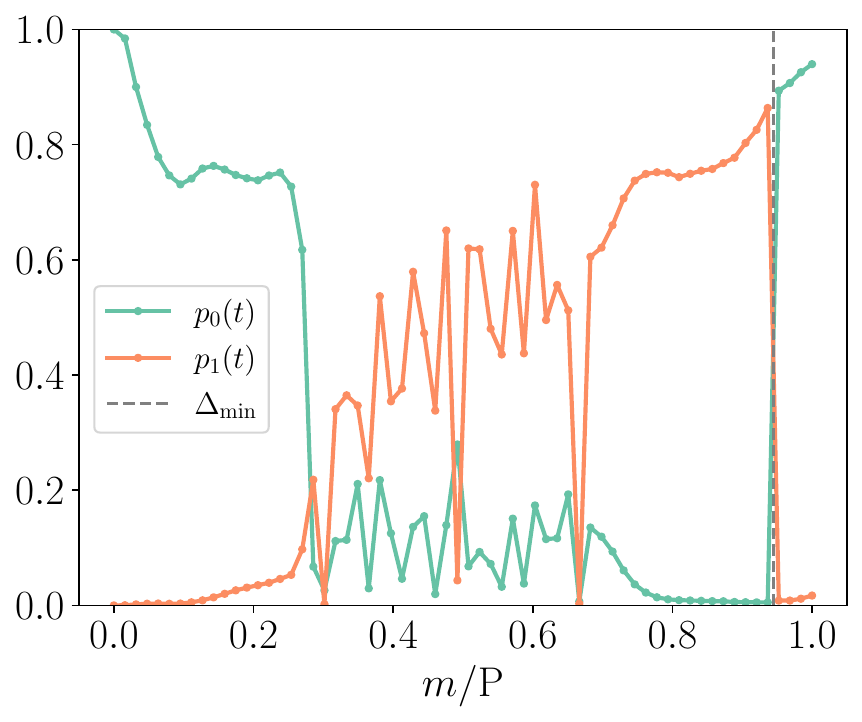}
\caption{(Top) The irregular optimal parameters for instance $1$ obtained by implementing dQA-F-CRAB for $\Ptrot = 64$ and $\Nbasis = 64$ without any warm start. We also plot the linear interpolation $\theta^x(t)$ and $\theta^z(t)$ utilized in the continuous-time dynamics, for which the x-axis should be intended as $t/\tau$. 
(Bottom) The discrete populations $p_j(m)$ for the digitized dynamics with irregular parameters. 
}
\label{fig:discrete_irregular}
\end{figure}
In this section, we show the benefits of working with smooth optimal parameters. To draw a comparison, we start by building an irregular set of optimal parameters for instance $1$. 
This is done by implementing dQA-F-CRAB for $\Ptrot = 64$ 
and $\Nbasis = \Ptrot$, 
without any warm-start for the optimization, i.e.,  
setting $\rmC_0 = 1$ and $\C^x=\C^z=0$ as initial conditions. 
The presence of high-frequency modes and the absence of a regularization induced by the iterative warm-start induce sudden variations, as a function of the layer index $m$, in the optimal parameters $\theta_m^x$ and $\theta_m^z$,
shown in the top panel of Fig.~\ref{fig:discrete_irregular}.
Despite its irregularity, such solution provides a value of residual energy $\epsilon^{\res}_{\scriptscriptstyle\mathrm{IRR}} = 0.00190$, very close to that of the smooth solution for $\Ptrot = 64$ and 
$\Nbasis = \Ptrot/2$, $\epsilon^{\res}_{\Ptrot} = 0.00183$, shown in the top-right panel of Fig.~\ref{fig:parameters_zhou}. 
In the digital framework, the smooth and the irregular solutions perform similarly, not only in terms of residual energy but also concerning the instantaneous populations of the first eigenstates. 
The bottom panel of Fig.~\ref{fig:discrete_irregular} shows that,
after a fast oscillatory phase, the population $p_1(t)$ of the first excited state becomes increasingly larger: the irregular solution performs similar STA dynamics, featuring the same population inversion described in the previous section for the smooth solution.

One of the key findings of our work is that only the smooth solutions can be used to define a valid continuous-time optimal control, which might be applied in a realistic experiment with real-time quantum dynamics. This procedure allows us to obtain an optimal QA schedule without the need for prior information on the minimum gap.
To show this, we consider the continuous-time evolution driven by the annealing Hamiltonian $\Ho(t)$
\begin{equation} \label{eq:annealing}
\Ho(t) = (1-s(t)) \, \Ho_x + s(t) \, \Ho_z \;, 
\end{equation}
where $s(t)=\theta^z(t)/(\theta^x(t)+\theta^z(t))$, 
with $\theta^x(t)$ and $\theta^z(t)$ obtained by interpolating the discrete optimal parameters $\theta^x_{m}$ and $\theta^z_{m}$.
This interpolation is shown in Fig.~\ref{fig:discrete_irregular}(top panel) for the irregular solution. 

\begin{figure}[htp]
\includegraphics[width=\columnwidth]{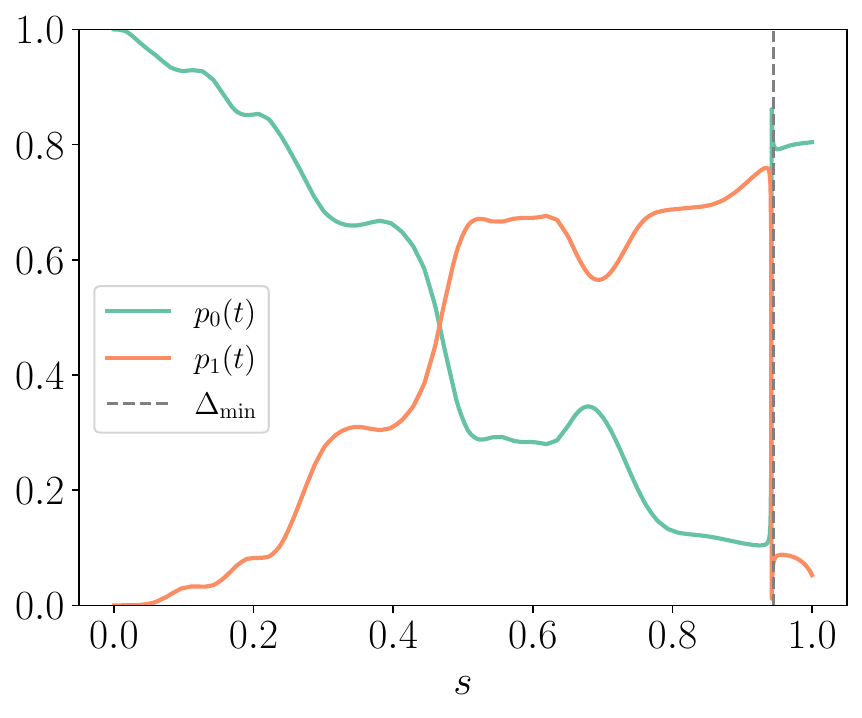}\\
\includegraphics[width=\columnwidth]{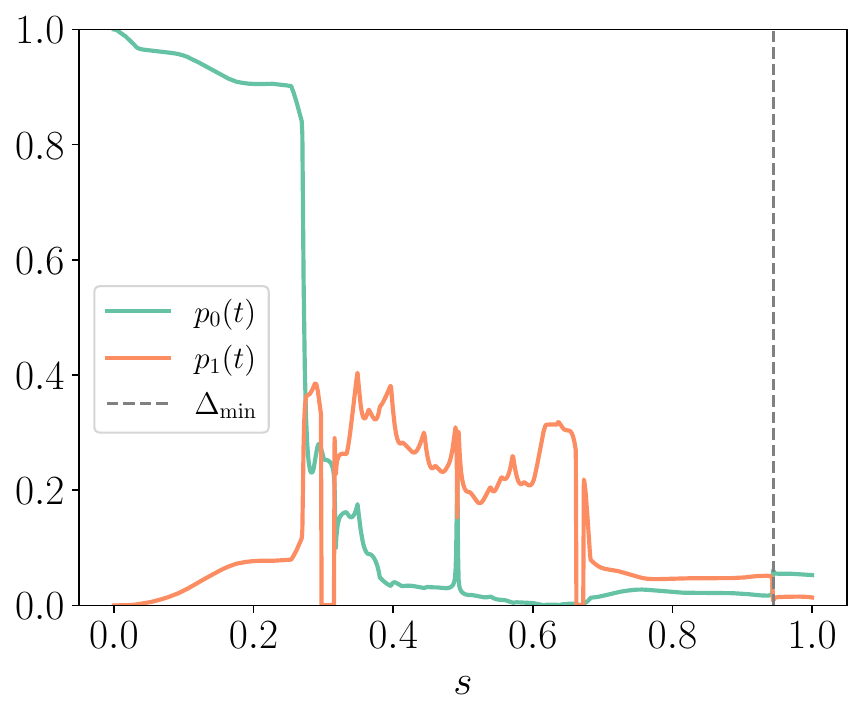}
\caption{(Top) Ground and first excited state populations $p_j(t)$ for the continuous-time evolution obtained by interpolation of the smooth digital parameters. (Bottom) The corresponding populations for the continuous-time evolution, as obtained by interpolation of the irregular digital parameters.   }
\label{fig:continuous_evolution}
\end{figure}
The state $|\Psi(t)\rangle$ is obtained by solving the Schr\"odinger 
equation \eqref{eqn:Schrodinger_eq}. 
To do so, we efficiently simulate the action of the full evolution operator using a scaling and squaring method, together with a truncated Taylor series approximation \cite{doi:10.1137/100788860}, which does not involve any Trotterization. 
In this continuous-time simulation, we used a sufficiently small 
integration time step ($dt_c = 0.1$ proved to be small enough) and 
we considered a total annealing time 
$\tau=\sum_{m=1}^{\Ptrot} (\theta^x_m+\theta^z_m)$~\cite{zhou_quantum_2020}. 
Finally, we compute the average energy 
$E^{\fin}(\tau)=\langle \Psi(\tau) | \Ho_z | \Psi(\tau)\rangle$, and the corresponding residual energy:
\begin{equation}
\epsilon^{\res}(\tau) = 
\frac{E^{\fin}(\tau) - E_{\min}}{E_{\max} -  E_{\min}} \;.
\end{equation}

Our simulations show that the continuous-time Schr\"odinger evolution 
with $\theta^x(t)$ and $\theta^z(t)$ obtained from the smooth optimal digital schedule $\theta^x_m$ and $\theta^z_m$ yields a low residual energy , $\epsilon^{\res}(\tau)=0.0078$.
On the contrary, when $\theta^x(t)$ and $\theta^z(t)$ are obtained from the irregular optimal $\theta^x_m$ and $\theta^z_m$ shown  
in Fig.~\ref{fig:discrete_irregular}(top panel), the result is very poor: $\epsilon^{\res}(\tau)=0.2382$.

This performance difference is also reflected in the population of the instantaneous eigenstates $p_j(t)$, shown in Fig.~\ref{fig:continuous_evolution}. 
We define the populations in the continuous-time framework as:
\begin{equation}
   p_j(t) = |\langle \phi^{\,j}(t) | \psi(t) \rangle |^2 , 
\end{equation}
where $\ket{\phi^{\,j}(t)}$ is the $j$-th eigenvector of $\Ho(t)$. 
For smooth parameters (top panel), despite some minor differences due to the time-discretization and Trotterization, the populations in the continuous-time evolution show qualitatively the same behavior as those in the digital evolution (see Fig.\ref{fig:populations_P64}): we retrieve the STA protocol, which succeeds in largely populating the target state at the end of the time evolution.

On the other hand, the continuous-time evolution induced by the irregular parameters yields very different population results: this can be observed by comparing the bottom panel of Fig.~\ref{fig:continuous_evolution} to  the bottom panel of Fig.~\ref{fig:discrete_irregular}, showing the populations for discrete dynamics.
Indeed, in the continuous-time evolution, the population of the low-energy spectrum drastically decreases after the irregular phase, and other excited states are significantly populated. 
As a consequence, the population inversion occurring at the minimum gap does not effectively populate the ground state. 

We conclude that sharp variations in the time-dependent Hamiltonian controls affect the continuous-time dynamics and result in a drastic loss of performance compared to the digital evolution. 
On the other hand, optimal digital smooth solutions are suitable to approach the problem also in continuous-time, 
and can possibly be refined by local optimization in continuous time.

\section{Conclusions}\label{sec:conclusions}
We have presented a set of methods for constructing optimal Quantum Annealing (QA) schedules for hard MaxCut problems where a small spectral gap is a bottleneck to standard adiabatic dynamics. 

The methods utilized fall into two distinct categories. The first consists of constructing a Trotter-digitized QA (dQA) 
with schedules parameterized in terms of Fourier modes or Chebyshev polynomials. These parameterizations are adapted from the Chopped Random Basis (CRAB) technique, which is employed for optimal control in continuous-time systems.
The second category is based on the Quantum Approximate Optimization Algorithm (QAOA)~\cite{Farhi_arXiv2014}, where solutions are found iteratively using linear interpolations (INTERP~\cite{zhou_quantum_2020} or LogINTERP~\cite{mbeng_quantum_2019}) or expansions in Fourier modes (FOURIER~\cite{zhou_quantum_2020}). 
In both cases, we have emphasized the importance of smooth optimal parameters and found that the same physical mechanism is behind the methods
that perform better: a ``\textit{shortcuts to adiabaticity}'' dynamics, 
with a population inversion occurring well before the minimum spectral gap is encountered. 

Summarizing the performance of the various methods on the hard instances we considered, we have found that QAOA-INTERP, 
based on iteratively increasing the layer depth at each step as $\Ptrot\to \Ptrot'=\Ptrot+1$, is not able to significantly outperform a linear-schedule dQA (dQA-LIN). 
On the contrary, its variant (QAOA-LogINTERP) based on duplicating the layer depth at each step, $\Ptrot\to \Ptrot'=2\Ptrot$, is among the best-performing methods.
QAOA-FOURIER~\cite{zhou_quantum_2020} also shows comparable performance, but its implementation is rather time-consuming.  
Both digitized CRAB approaches, namely dQA-F-CRAB and dQA-C-CRAB, perform equally well. The latter, in particular, requires the least computational resources to attain smooth optimal solutions: 
for given $\Ptrot$ and $\Nbasis$, dQA-C-CRAB finds optimal and smooth solutions using a discrete linear schedule as the initial condition without any iterative warm start.

A noteworthy feature of the smooth optimal solutions is their transferability among hard instances of MaxCut. Indeed, we find that the solutions found for a given instance perform rather well, even without any further local optimization, on other hard instances. If further local optimization is applied, the transferred solution proves even better than what you would obtain by directly optimizing the target instance.

Finally, we find that smooth optimal protocols are a good starting point for a time-continuous schedule to be implemented in an analog device.  
Remarkably, this provides an optimized QA protocol that improves on the standard linear schedule, without requiring preliminary information on the instantaneous gap.

Regarding future work, we identify two promising avenues for exploration.
The first avenue involves applying the techniques discussed to one of the simplest models featuring an exponentially small spectral gap in its spectrum: the frustrated Ising ring model, as introduced in \cite{Knysh_PRA2020} and further explored in \cite{Côté_2023}.
The second direction is to further explore the connection between our digitized solution and the ``shortcut to adiabaticity'' mechanism behind a continuous-time approach based on counter-diabatic driving~\cite{KOLODRUBETZ20171}. 
Recent work~\cite{Wurtz_2022} has pointed out that the first-order correction in the Baker-Campbell-Hausdorff formula has exactly the same structure and sign as an adiabatic gauge potential term realizing a counter-diabatic driving: we plan to explore this connection between digitized and continuous-time dynamics further in the near future.

\begin{acknowledgments}
We acknowledge Valentina Ros for very useful discussions.
G.E.S. acknowledges financial support from PNRR MUR project PE0000023-NQSTI, from PRIN 2022H77XB7 of the Italian Ministry of University and Research, and 
from European Union's H2020 Framework Programme/ERC Advanced Grant N. 8344023 ULTRADISS. 
GES and GP acknowledge financial support from the QuantERA II Programme STAQS project that 
has received funding from the European Union’s H2020 research and innovation programme under Grant Agreement No 101017733.
GES acknowledges that his research has been conducted within the framework of the Trieste Institute for Theoretical Quantum Technologies (TQT).
Views and opinions expressed are however those of the author(s) only. 
Neither the European Union nor the granting authority can be held responsible for them.
\end{acknowledgments}

\appendix

\begin{figure*}[htp]
\captionsetup[subfigure]{labelformat=empty}
\subfloat[]{%
  \includegraphics[width=\columnwidth]{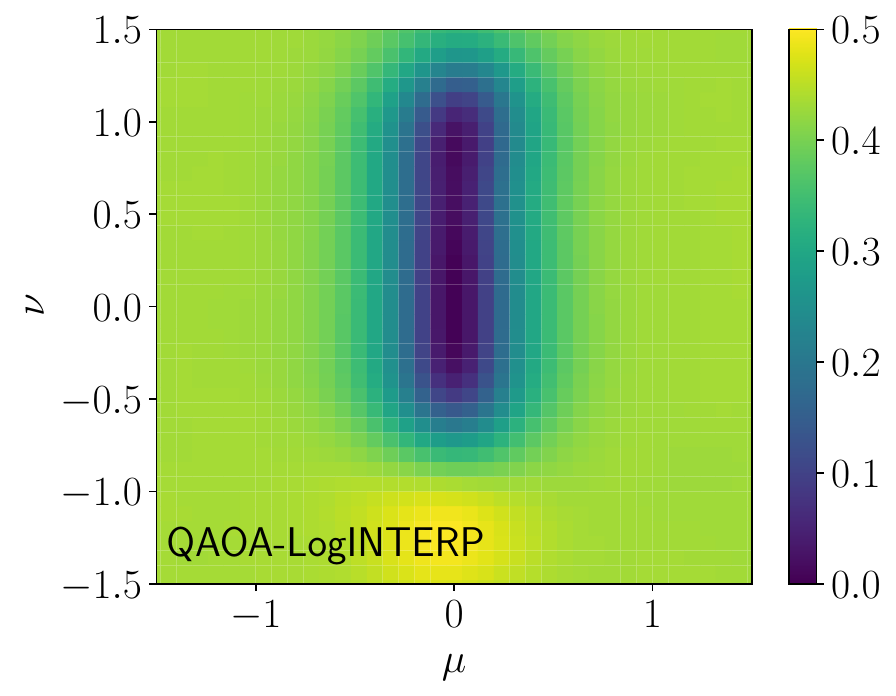}%
  \includegraphics[width=\columnwidth]{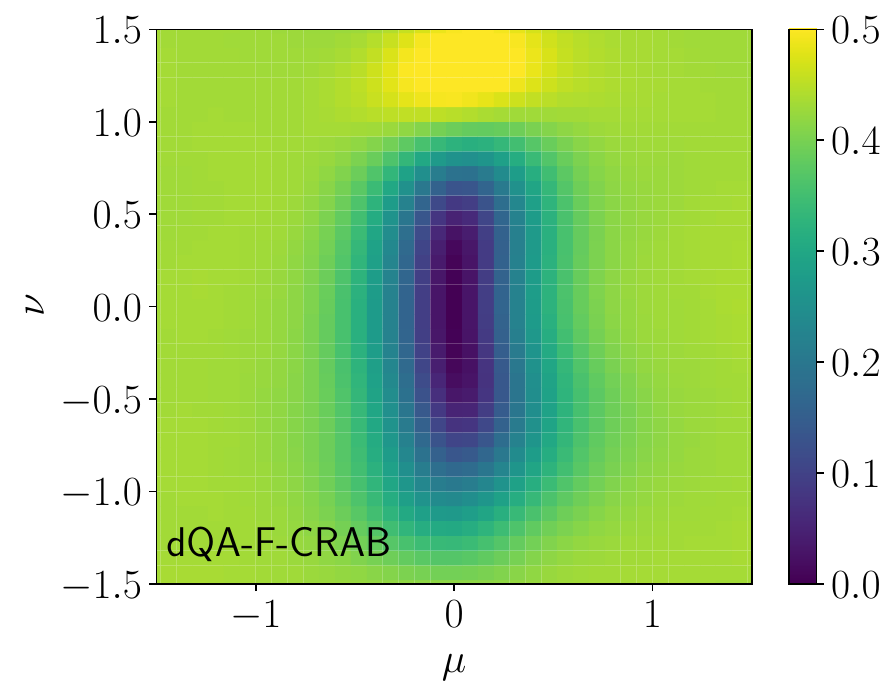}%
}
\vspace{-9mm}
\subfloat[]{%
  \includegraphics[width=\columnwidth]{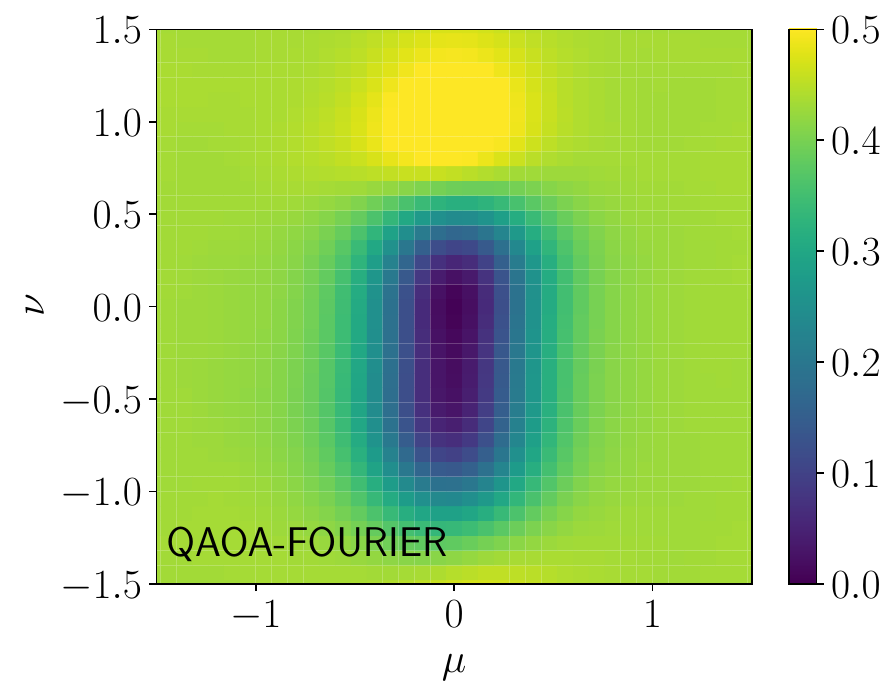}%
\includegraphics[width=\columnwidth]{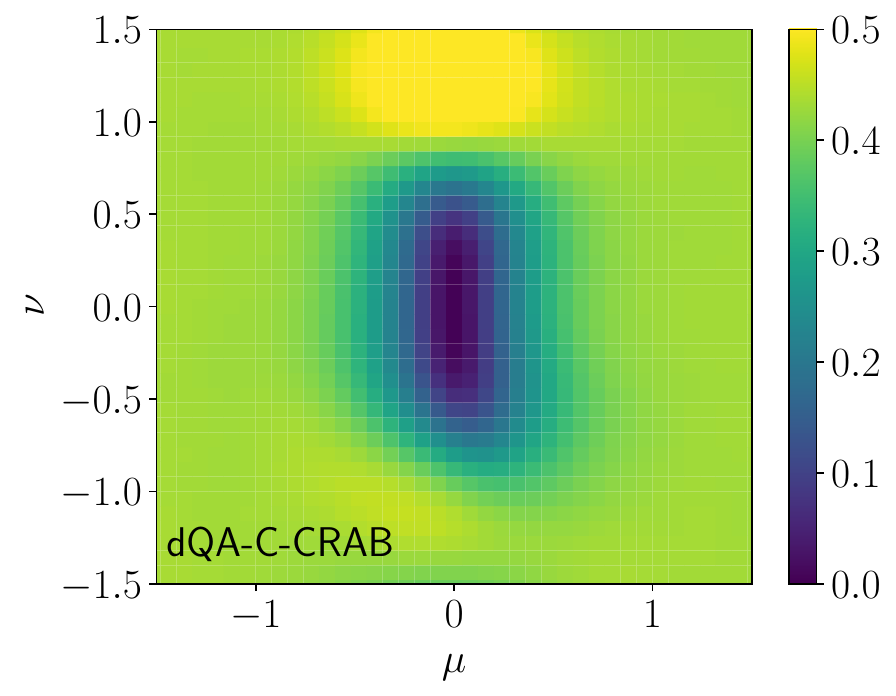}%
}
\vspace{-7mm}
\caption{The residual energy $\epsilon^{\res}_{\Ptrot}(\btheta)$ for different methods along two directions in parameter space, $\mu$ and $\nu$. These correspond, respectively, to the direction of maximum and minimum curvature of $\epsilon^{\res}_{\Ptrot}(\btheta)$ around the corresponding local minimum (see Fig.\ref{fig:parameters_zhou}), set at $\mu = \nu = 0$.}
\label{fig:hessians}
\end{figure*}  

\section{Gradient-based optimization} \label{app:gradient}
Our classical simulations rely on BFGS as a gradient-based algorithm for local optimizations.
To speed up the simulation runs and to avoid numerical approximations of the gradient,
we calculate the analytic expression for the gradient of the variational energy in Eq.~\ref{eqn:E_dQA}, exploiting
the digitized form of the variational state. Let $\rmC_j$ denote a generic parameter in the list $\C=(\rmC_0,\C^x,\C^z)$.
Hence:
\begin{eqnarray}
\frac{\partial E_{\Ptrot}^{\dQA}}{\partial \rmC_j} &=&  
\langle \Psi_{\Ptrot}^{\dQA} | \Ho_z {\textstyle | \frac{\partial \Psi_{\Ptrot}^{\dQA}}{\partial \rmC_j} \rangle} + 
{\textstyle \langle \frac{\partial \Psi_{\Ptrot}^{\dQA}}{\partial \rmC_j}} | \Ho_z |\Psi_{\Ptrot}^{\dQA}  \rangle \nonumber \\
&=& 2 \, \Real \langle \Psi_{\Ptrot}^{\dQA} | \Ho_z {\textstyle | \frac{\partial \Psi_{\Ptrot}^{\dQA}}{\partial \rmC_j} \rangle}  \;. 
\end{eqnarray}
The derivative of the state follows from the chain rule. Let us denote by $\theta^{\alpha}_m$ the parameters, with $\alpha=x,z$. 
Then:
\begin{equation}
{\textstyle | \frac{\partial \Psi_{\Ptrot}^{\dQA}}{\partial \rmC_j} \rangle} = \sum_{\alpha=x,z} \sum_{m=1}^{\Ptrot} 
{\textstyle | \frac{\partial \Psi_{\Ptrot}^{\dQA}}{\partial \theta^{\alpha}_m} \rangle} {\textstyle \frac{\partial \theta^{\alpha}_m}{\partial \rmC_j}} \;.
\end{equation}
The two ingredients appearing here are simple to calculate from Eq.~\eqref{eq:theta_x_z}:
\begin{equation}
\left\{
\begin{split}
    \frac{\partial \theta^{x}_m}{\partial \rmC_0} &= 
\frac{\Ptrot-(m-\frac{1}{2})}{\Ptrot} \vspace{3mm} \\
\frac{\partial \theta^{z}_m}{\partial \rmC_0} &= 
\frac{m-\frac{1}{2}}{\Ptrot} \vspace{3mm} \\
\frac{\partial \theta^{x}_m}{\partial \rmC^x_n} &= \frac{\Ptrot-(m-\frac{1}{2})}{\Ptrot} f_n(t_m) \vspace{3mm} \\ 
\frac{\partial \theta^{z}_m}{\partial C^z_n} &= \frac{m-\frac{1}{2}}{\Ptrot} f_n(t_m)
\end{split}
\right. \;.
\end{equation}
The other ingredient, the state derivative, depends on the Trotter splitting performed. 
For simplicity, let us focus on the lowest-order Trotter splitting. 
Define, for $m=1\cdots\Ptrot$, the shorthands:
\[ \Uo_m = \Uo(\theta^x_m,\theta^z_m) = \nep^{-i\theta^x_m\Ho_x} \, \nep^{-i\theta^z_m\Ho_z} \;, \] 
and 
\begin{equation}
|\Psi_m\rangle = \Uo_m \cdots \Uo_1 |\Psi_0\rangle \;.     
\end{equation}
\begin{widetext}
Then: 
\begin{equation}
i \langle \Psi_{\Ptrot}^{\dQA} | \Ho_z {\textstyle | \frac{\partial \Psi_{\Ptrot}^{\dQA}}{\partial \theta^x_m} \rangle} 
=  \langle \Psi_m| \Uo^{\dagger}_{m+1} \cdots  \Uo^{\dagger}_{\Ptrot} \Ho_z \Uo^{\phantom \dagger}_{\Ptrot} \cdots \Uo^{\phantom \dagger}_{m+1} \Ho_x | \Psi_m\rangle \;,
\end{equation}
and
\begin{equation}
i \langle \Psi_{\Ptrot}^{\dQA} | \Ho_z {\textstyle | \frac{\partial
\Psi_{\Ptrot}^{\dQA}}{\partial \theta^z_m} \rangle} 
= \langle \Psi_{m-1}| \Uo^{\dagger}_{m} \cdots  \Uo^{\dagger}_{\Ptrot} \Ho_z \Uo^{\phantom \dagger}_{\Ptrot} \cdots \Uo^{\phantom \dagger}_{m} \Ho_z | \Psi_{m-1}\rangle \;.
\end{equation}
\end{widetext}
The complete gradient can thus be calculated efficiently in a classical simulation by storing intermediate states $|\Psi_m\rangle$ and combining them to compute each partial derivative.

\begin{figure}[htp]
\captionsetup[subfigure]{labelformat=empty}
  \includegraphics[width=\columnwidth]{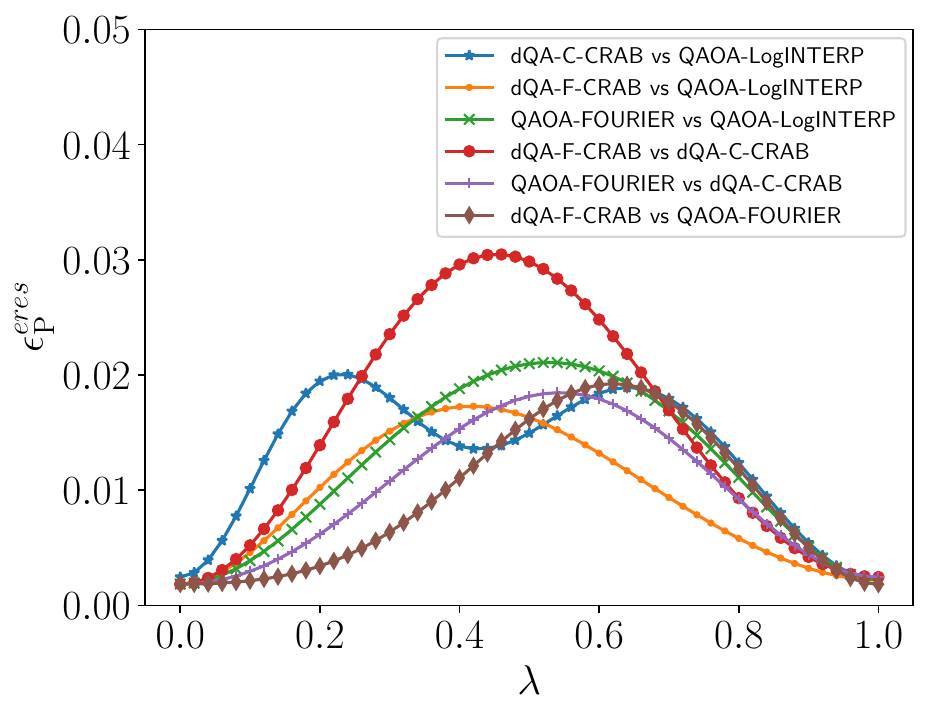}%

\caption{The residual energy $\epsilon^{\res}_{\Ptrot}(\btheta^{12}(\lambda))$ along the direction $\lambda$ linking two local minima in parameter space. The corresponding optimal parameters are shown in Fig.\ref{fig:parameters_zhou}}
\label{fig:interpolation_params}
\end{figure}  

\section{The FOURIER approach of Zhou {\em et al.}}
\label{app:FOURIER}
We give here a few details on the implementation of the FOURIER method proposed in Ref.~\cite{zhou_quantum_2020}, reformulated in a notation consistent with the previous sections.

\textbf{a) $\Nbasis=\Ptrot$, without randomness.}
In this case, one should keep the number of Fourier terms $\Nbasis$ equal to $\Ptrot$ and hence increase both by $1$ at each step~\cite{zhou_quantum_2020}.
Suppose to have the optimal --- most likely, {\em locally} optimal --- Fourier coefficients $\C^{(*,\Ptrot-1)}$ at level $\Ptrot-1$. 
Then, take as an initial guess, for both $\C^z$ and $\C^x$:
\begin{equation} \label{eqn:candidate_a}
    \C^{(0,\Ptrot)} = (\C^{(*,\Ptrot-1)},0) \;, 
\end{equation}
and determine the new minimum $\C^{(*,\Ptrot)}$ by a local search strategy starting from $\C^{(0,\Ptrot)}$.

\textbf{b) $\Nbasis=\Ptrot$, with randomness.}
As in \textbf{a)} above, consider a (local) optimal solution $\C^{(*,\Ptrot-1)}$ and construct a candidate initial point as in Eq.~\eqref{eqn:candidate_a},
with a corresponding new mininum $\C^{(*,\Ptrot)}$.
On top of that, multiple initial points are randomly generated (see below), starting from the {\em best} final solution $\C^{(\best,\Ptrot-1)}$ kept at each $\Ptrot$: 
notice that $\C^{(\best,\Ptrot-1)}$ might differ from $\C^{(*,\Ptrot-1)}$. 
More precisely, starting from  $\C^{(\best,\Ptrot-1)}$, one should generate $R>0$ random candidate initial points, e.g., typically, $R=10$,
as follows:
\begin{equation}
    \C^{(0,\Ptrot)} = (\C^{(\best,\Ptrot-1)}+\alpha \, \mathbf{r},0) \;.
\end{equation} 
Here, $\alpha$ is a real coefficient, typically set to $0.6$, and $\mathbf{r}$ is a $(\Ptrot-1)$-dimensional vector of gaussian-distributed random numbers
with zero mean and variance given by $\mathrm{r}_n =  (\rmC^{(*,\Ptrot-1)}_n)^2$. 
An extra initial point is generated with $\mathbf{r}=0$. 
Correspondingly, candidate final minima are found with a local optimization starting from each of these $R+1$ initial points. 
The new $\C^{(\best,\Ptrot)}$ at level $\Ptrot$ is selected as the best solution in a set of $R+2$ candidates comprising the $R+1$ minima just found, and 
the $\C^{(*,\Ptrot)}$ generated as in \textbf{a)}.
Note that, for the next iterative step, both $\C^{(*,\Ptrot)}$ and $\C^{(\best,\Ptrot)}$ need to be saved.

\textbf{c) $\Nbasis$ fixed, with randomness.}
Everything is performed as in case \textbf{b)} above, only up to the value of $\Ptrot=\Nbasis$. Beyond that, $\Nbasis$ stays unchanged while $\Ptrot$ is further increased up to the
desired $\Ptrot_{\max}$.

\begin{figure*}[htp]
\captionsetup[subfigure]{labelformat=empty}
\includegraphics[width=0.3333\textwidth]{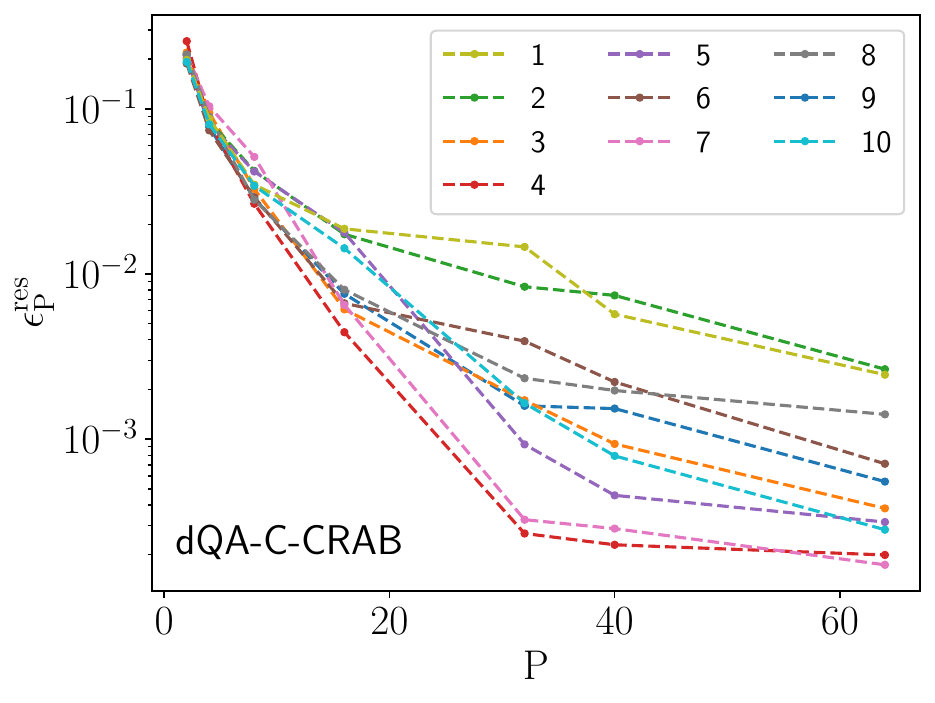}%
\includegraphics[width=0.3333\textwidth]{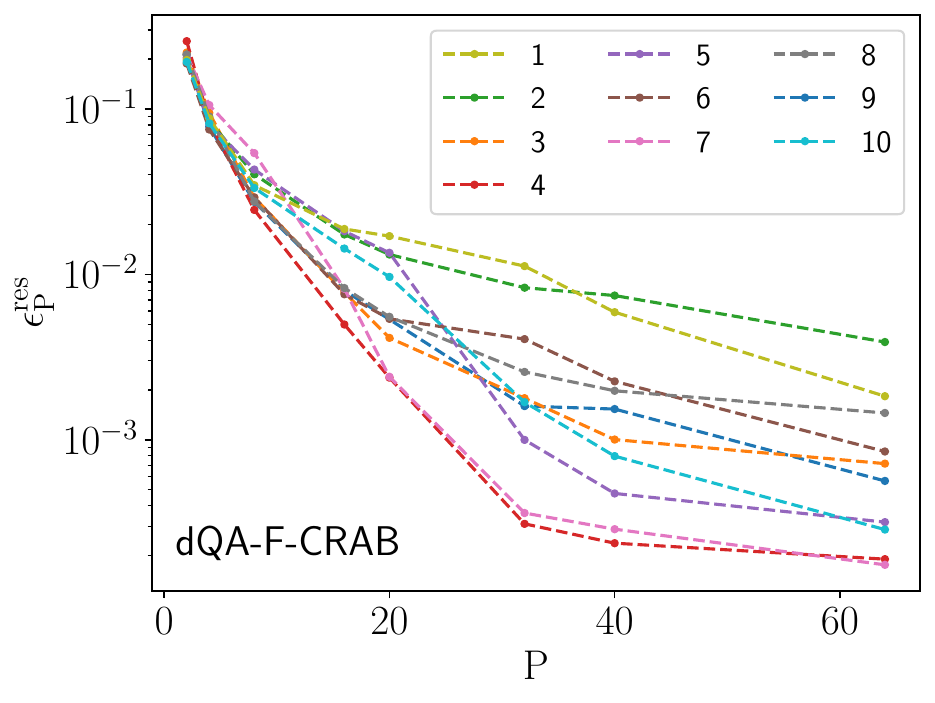}%
\includegraphics[width=0.3333\textwidth]{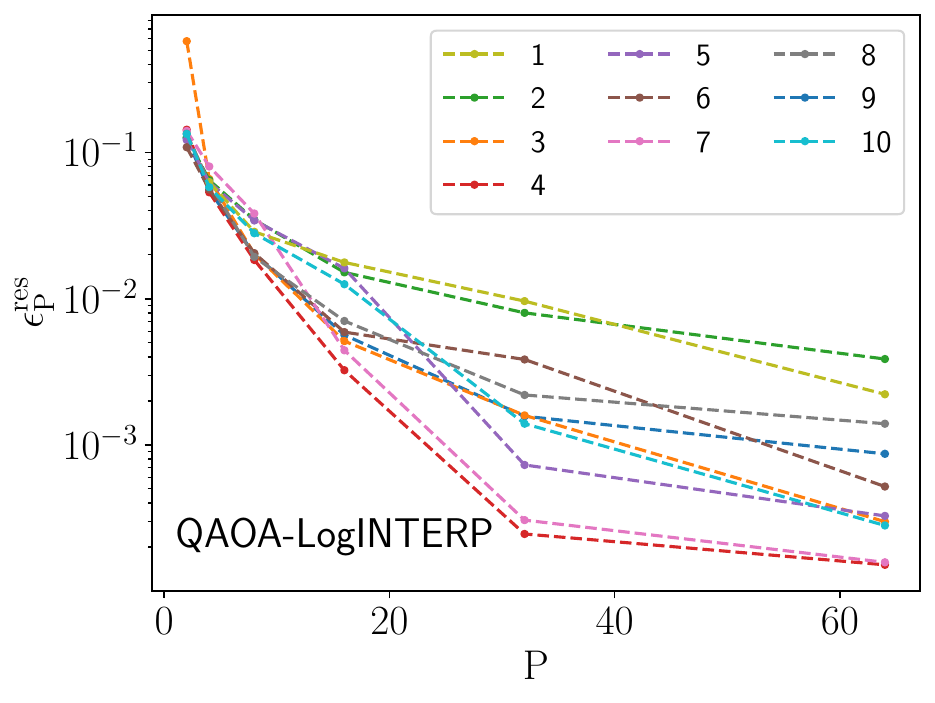}%

\caption{Residual energy as a function of $\Ptrot$ for the $10$ hard instances we considered and for different methods. For each instance, these methods provide similar final results. } 
\label{fig:eres_other_instances}
\end{figure*}

\section{Local neighborhood of different optimal solutions} \label{app:Hessian}
As the high complexity of the parameter landscape might hinder some connections among seemingly different local minima, in the following we further investigate this issue.
Starting from the behavior of the residual energy in the neighborhood of these local minima, our goal is to determine whether the solutions we find are actually distinct in the $2\Ptrot$-dimensional parameter space. In particular, we focus on the Hessian of the residual energy $\epsilon^{\res} (\btheta)$ as a function of the variational parameters $\btheta$. 
For each method described in Section~\ref{sec:methods}, we evaluate the Hessian of the energy in the minimum $\epsilon^{\res} (\btheta^*)$ using finite differences on top of the analytical expression of the first-order gradient. The eigenvectors of the Hessian characterize the local curvature of the residual energy: the corresponding eigenvalues allow us to determine the directions where the curvature is locally maximum and minimum. In Fig.\ref{fig:hessians}, we plot $\epsilon^{\res} (\btheta)$ along these two directions. Specifically, we plot the residual energy as a function of the displacements $\mu$ and $\nu$ along the direction of maximum and minimum curvature, respectively. The displacements are evaluated from $\epsilon^{\res} (\btheta^*)$, which is therefore located in the origin $\mu = \nu = 0$. 
We see that, even in the direction of minimum local curvature $\nu$, a variation in the parameter $\btheta$ results in a sizable variation of the residual energy.
Intuitively, the four solutions determined by the different methods appear to be isolated minima in the $2\Ptrot$-dimensional parameter space.

Furthermore, for each pair of local minima obtained through different methods, we focus on their convex linear combination.
More in detail, given two points in parameter space $\btheta^{(1)}$ and $\btheta^{(2)}$, we consider:
\begin{equation}  
  \btheta^{(12)}(\lambda) = \lambda \ \btheta^{(1)} + (1- \lambda) \ \btheta^{(2)}
\end{equation}
and we compute the corresponding residual energy as a function of the interpolating parameter $\lambda \in [0,1]$. 
We show the results in Fig.\ref{fig:interpolation_params}: each pair of optimal solutions is separated by a barrier where the residual energy increases by an order of magnitude. This finding precludes the existence of a spherical broad minimum in the variational energy landscape where two of these optimal local minima rely. However, we stress that a complicated path in parameter space could connect two local minima without necessarily crossing a residual energy barrier.  

\section{Performance over different instances} 
\label{app:other_instances}
In this section, we summarize the performance of the methods presented in the main text on the $10$ hard instances we considered. In Fig. \ref{fig:eres_other_instances}, we show the residual energies for dQA-C-CRAB, dQA-F-CRAB, and QAOA-LogINTERP. Consistently with the results shown in the main text, these methods implement ``shortcuts to adiabaticity'' protocols and outperform the linear dQA-LIN method. 
For each instance in exam, dQA-C-CRAB, dQA-F-CRAB, and QAOA-LogINTERP consistently succeed in decreasing residual energy as the circuit depth $\Ptrot$ increases. 
Empirically, a linear interpolation method, QAOA-INTERP, often performs significantly worse.
Finally, as expected, the residual energy of dQA-LIN reaches a plateau, stabilizing at a constant value as $\Ptrot$ increases.

\bibliography{Biblio_QOC_MaxCut.bib}
\end{document}